%% file: sksndirection.tex
\documentclass{aastex631}

\usepackage{derivative}
\usepackage{amsmath}
\usepackage{etoolbox}
\makeatletter
\patchcmd{\ltx@foottext}{%
  .5\textwidth\advance\hsize-18pt}{%
  \linewidth\advance\hsize-1.8em%
  }{}{}
\makeatother

\newcommand{\degree}{$^{\circ}$}
\newcommand{\dtheta}[1]{$\theta_{SN}^{~#1\%}$}
\newcommand{\avgdtheta}{$\theta_{SN}^{~avg}$} 
\newcommand{\cthsn}{$\cos\theta_{SN}$}
\newcommand{\thsn}{$\theta_{SN}$}

\newcommand{\SKbig}{Super-Kamiokande~}
\newcommand{\snw}{SNWATCH}
\newcommand{\snww}{SNWATCH~}
\newcommand{\hp}{HEALPix~}
\newcommand{\oxccnue}{O16CC1}
\newcommand{\oxccnuebar}{O16CC2}
\newcommand{\esnue}{ES1}
\newcommand{\esnuebar}{ES2}
\newcommand{\esnux}{ES3}
\newcommand{\esnuxbar}{ES4}
\newcommand{\ns}{NSIDE}

\newcommand{\dsntrue}{$\hat{d}_{sn}$}
\newcommand{\dsnrecon}{$\hat{d}_{sn}^{~recon}$}
\newcommand{\dnusn}{$\hat{d}_{sn{\text -}\nu}$}

\newcommand{\dnusnrecon}{$\hat{d}_{sn{\text -}\nu}^{~recon}$}

\begin{document}
% \linenumbers

\title{Development of Faster and More Accurate Supernova Localization at Super-Kamiokande}

\input{authors-merge-barry-orcid2.tex}

\correspondingauthor{Barry W. Pointon}
\email{barry\_pointon@bcit.ca}

\begin{abstract}
The next nearby core-collapse supernova (SN) promises to yield a treasure of scientific information through multi-messenger astronomy. Early observations of the shock breakout (SBO) emissions are especially critical to understand the SN explosive mechanism as well as the properties of the progenitor star. Neutrino observatories are able to provide an early alert of a SN before the arrival of the SBO radiation.
Super-Kamiokande (SK) has the unique capability to independently reconstruct an accurate SN pointing direction as part of its real-time monitoring system, ``\snw.'' Recent upgrades to SK by adding gadolinium (Gd) to the detection volume have been accompanied by efforts to improve the speed and accuracy of SN direction reconstruction. 
A new, novel HEALPix-based approach (``HP-Fitter'') can calculate the SN direction from the reconstructed burst event directions in less than one second. As well, the previous maximum-likelihood direction fitter (``ML-Fitter'') was upgraded by incorporating event information from Gd neutron-capture as well as using the HP-Fitter for the initial fit parameters and from code refactoring and optimization. The improved ML-Fitter has better angular resolution but direction reconstruction time is 
$\mathcal{O}$(sec). Together with improvements in burst detection and event reconstruction times, \snww is now able to generate an SN alert with pointing information in about 90 seconds. These upgrades have been implemented at SK and integrated into a new automated system to provide GCN notices. 
\end{abstract}

%TC:ignore


\section{Introduction} \label{sec:intro}
\subsection{The Next Galactic Supernova} \label{subsec:next_sn}
The next galactic core-collapse supernova (SN) is highly anticipated by both the world's scientific community and the interested public~\citep{Castel2022}. 
The most recent nearby event, SN1987A, validated the basic features of theoretical SN models, revealed new directions for research and continues to be the subject of intense study. 
The detection of neutrinos from SN1987A was a milestone in multi-messenger astronomy~\citep{PhysRevLett.58.1490, PhysRevLett.58.1494}. 

Today, a diverse array of advanced, multi-wavelength and multi-messenger instruments is ready to observe the next nearby SN in unprecedented detail.
Included in these ranks are dedicated neutrino detectors and other neutrino-sensitive experiments that can characterize the time, energy and neutrino-flavor distributions in the SN neutrino burst
\citep{Scholberg:2018meq}. 
SN neutrino burst detection in an important physics goal for many experiments, including Super-Kamiokande (SK) \citep{Abe:2016waf, Kashiwagi_2024}, IceCube~\citep{refId0, Abbasi:20239F}, DUNE~\citep{Abi:2021}, JUNO \citep{Abusleme2023}, KM3NET\citep{Aiello2022}
and several experiments at SNOLAB \citep{caden2025supernovadetectionsnolab}.
In addition, these detectors are able to provide an early warning alert for other observers. Of these, \SKbig (SK) also has the unique capability to independently determine the SN direction and provide accurate pointing information from the neutrino burst events. 
The opportunity afforded by an early warning with location information allows multi-messenger observatories to prepare individual SN alert response plans and the scientific community as a whole to make coordinated, global strategies. 

The rarity of galactic SNe gives urgency to such preparations.
Searches for SN bursts by long-term neutrino experiments give upper limits of $<$ 9~{(100~yr)}$^{-1}$ out to 25 kpc for LVD based on a 21 year search~\citep{Agafonova_2015} and $<$29~{(100~yr)}$^{-1}$ out to 100 kpc for SK based on a ten year search~\citep{Mori:2022aaa}.
A recent global study puts the estimated rate at $1.63 \pm 0.46$ {(100~yr)}$^{-1}$
with a mean time between events of $61\substack{+24 \\ -14}$\ yr~\citep{ROZWADOWSKA2021101498}.

\subsection{SN Neutrinos and EM Emissions} \label{subsec:sn_nu_det}
Theoretical SN models predict that $>$99\% of the energy released by core collapse is converted into a burst of $\mathcal{O}(10^{58})$ neutrinos~\citep{RevModPhys.74.1015, Fischer:2008rh, Mirizzi:2015eza}.
These play a significant, if poorly understood, role in the explosion mechanism. However, this enormous luminosity enables the detection of SN burst neutrinos from large distances, as corroborated by the detection of neutrinos from SN1987A at $\sim$50 kpc~\citep{PhysRevLett.58.1490}. 
% Detailed observations of the SN neutrino burst are expected to provide vital information
% on the progenitor star, the supernova explosion mechanism, neutrino-matter interactions, such as flavor-changing oscillations, neutrino-neutrino interactions and collective excitations at high density (such as SASI oscillations)
% \citep{totani, Scholberg:2018meq, Adams:2013ana, Bendahman_2024}
% As well, these observations may yield information on neutrino properties, such as the ordering of the mass hierarchy, \citep{Horiuchi:2017sku}. 
% Therefore, many neutrino-sensitive experiments, such as SK \citep{Abe:2016waf}, IceCube~\citep{refId0, Abbasi:20239F}, Dune~\citep{Abi:2021}, and Juno \citep{Abusleme2023}, have active programs to support the observation of a SN burst,   
Neutrinos are produced by various processes proximate to the time of core collapse. 
Although, they are initially trapped due to the extreme densities in the local environment, the outflow of matter following rebound quickly makes the core transparent to neutrinos, allowing them to stream freely through the disrupted stellar envelope after $\mathcal{O}$(msec). 
Thus, neutrinos, along with gravitational waves \citep{KOTAKE2013318}, would be first astronomical messengers of this cataclysmic event. 
Large numbers of neutrinos are also produced in accretion and cooling processes, but emissions drop rapidly after $\sim$20 seconds following core collapse.
The resulting neutrino fluence at earth depends only on the total luminosity and the SN distance, but would be sufficient for larger neutrino detectors to identify a burst from SNe anywhere in, or close to, the Milky Way. 
This provides the opportunity for making an SN early warning alert system.

The earliest EM emissions are from the shock breakout (SBO) \citep{Waxman2017}. As the SN shock front moves through the stellar envelope it slows
and near the edge of the star the density of the overlying stellar material decreases. 
When the front encounters material with an optical depth $\sim c/v$, SBO occurs, producing a short flash of soft X-rays/ultraviolet(UV), lasting seconds to minutes. 
This is followed by the emission of ultraviolet/optical radiation with a duration of seconds to days as the expanding stellar envelope cools after being disrupted by the breakout front. This is followed by the well-known plateau light emissions that may last several months. 

Measurements of the luminosity, spectral distribution and duration of the SBO flash and cooling emissions would provide unique information on the progenitor star; its radius, surface composition and mass-loss history.
The time delay between the neutrino burst and the SBO (``shock propagation time'') also depends on the state of the progenitor star. Although SBO generally occurs near the edge of the star, if substantial circumstellar material was accumulated from earlier winds or material outbursts, SBO will occur at a great radius which increases the shock propagation time. 
\citet{Kistler:2013aa} describes how the shock propagation time and the SBO duration may be used to reconstruct a ``tomographic'' snapshot of the progenitor star at the moment of the core collapse. 
For high-temperature, late-stage stars with significant envelope loss, such as Wolf-Rayet stars, the shock propagation time is $\mathcal{O}$(min) and the SBO duration is $\mathcal{O}$(sec). For Red Supergiants with a substantial stellar envelope remaining, the propagation time may be on the order of days with SBO duration $\mathcal{O}$(hr).

The magnitude of SN emissions at a given wavelength depends on the SN luminosity, distance and extinction by interstellar dust.
\cite{Nakamura:2016kkl}, following \cite{Adams:2013ana}, investigated the overall probability of detecting the plateau and SBO optical emissions for SNe at 8.5 kpc using randomized SN locations with simple models for the galactic distributions of SNe and interstellar dust. 
Their results are that for $\sim$23.5\% of SNe, the plateau optical emissions would be $>$ 25 mag and could not be detected by any current instrument. About $\sim$9.4\% of SNe would be at $\sim$20--25 mag, and could only best observed by the largest telescopes ($\sim$4--8 m class). 
These generally have detectors with a small field of view (FOV), the largest (3.5 deg, 9.6 deg$^2$) is for the 8.4-meter, Simonyi Survey Telescope equipped with the LSST Camera, at the Vera Rubin Observatory. 
The other $\sim$67.1\% of SNe would have optical magnitude $<$ 20 mag, and should be detectable by large FOV survey telescopes, except when limited by saturation.
The plateau emissions in the near infrared (NIR) would be less impacted by extinction, and expected to be at $\sim$$-2$ $\pm$ 2 mag. 
For SBO radiation, the optical magnitudes would be about 1 mag lower than the plateau and the NIR magnitudes would about 2 mag lower than the plateau NIR.
In the UV, where the SBO emissions peak, the extinction is severe, so the UV might only be detected if the SN is in opposition to the galactic center. 

\subsection{The Need for Fast and Accurate SN Localization Information} \label{subsec:sn_followup}
An early warning SN alert based on neutrino burst detection can be generated by most neutrino detectors, either independently, such as SK's real-time monitoring system, (SNWATCH), or as part of a coordinated network, such as the ``SuperNova Early Warning System'', (SNEWS)~\citep{Al_Kharusi:2021}. 
While a basic alert can inform observers of the impeding arrival of EM emissions, \cite{Nakamura:2016kkl} emphasizes the need for accurate SN pointing information to improve the opportunities to detect and study the EM emissions. 
Currently, SK is the only neutrino detector able to independently determine the SN direction from burst event data and provide accurate pointing information.

In the event of the next Galactic SN, such pointing information would be crucial if a SN is too faint to be found by a survey telescope, but is within the limiting magnitude of a larger telescope. Without pointing information the SN might go completely unobserved. 
Since the detector FOV on large telescopes is small, the efficiency of targeting the SN will depend on the accuracy of the pointing information.
If the SN is too faint for even the largest instruments, the pointing information may be useful in identifying the progenitor star in 
earlier surveys.

Of even greater value would be observations of the earliest SBO emissions, and determining the SBO light arrival time. 
This would require that an instrument be prepared and oriented so that the SN is within the detector FOV as close to the light arrival time as possible. 
The likelihood of such an opportunity depends on the size of the detector FOV relative to the accuracy of the pointing information. 
It also requires the alert to be generated and the instrument prepared and slewed before the light arrival.
As described in \citet{Kistler:2013aa}, the shock propagation time, and therefore the time between the arrival of the neutrinos and the SBO radiation varies with the progenitor, from seconds to hours. 
Therefore, the speed with which an SN alert is generated (``alert latency'') could impact how early the SN light can be observed.

SNWATCH has been operating for nearly a decade. Recent upgrades to the SK detector have been accompanied by improvements to SNWATCH that increase the sensitivity for detecting bursts, reduce event reconstruction time, and improve the speed and accuracy of the SN direction reconstruction~\citep{Kashiwagi_2024}. This has reduced the alert latency and increased the accuracy of the pointing information. 

This paper specifically describes the improvements to the SNWATCH SN direction reconstruction methods, which led to significant gains in speed and accuracy. Section~\ref{sec:skburstdetection} discusses SN burst neutrino detection at SK, including the impact of the SK-Gd upgrades. In Section~\ref{sec:healpixfitter} a new, novel SN direction reconstruction method based on HEALPix (the ``HP-Fitter'') is described. Section~\ref{sec:mlfitterimprov} details the improvements to the standard maximum-likelihood based, SN direction fitter (the ``ML-Fitter''). The methods used to test and optimize the performance of the new and updated fitters are explained in Section~\ref{sec:methods} with the results and discussion in Section~\ref{sec:results}. Finally, Section~\ref{sec:summary} gives a summary of these results and a discussion of how these developments might prompt revision of SN alert follow-up plans by observers. 

\section{Supernova Burst Monitoring at SK} \label{sec:skburstdetection}
\subsection{The Super-Kamiokande Detector} \label{subsec:superk}
SK is the premier large volume, water Cherenkov neutrino detector,
located 1000 m below the top of Mount Ikeno, in Gifu prefecture, Japan~\citep{Fukuda:2002uc}. 
The external rock overburden ($\sim$2.7 km water equivalent) also provides substantial passive shielding. 
The detector consists of 50 kilotons of water in a cylindrical stainless steel tank. 
The water volume is divided by an internal concentric photosensor mounting structure into a large inner detector (ID), the primary detection volume, and a smaller volume outer detector (OD) that acts as active and passive shielding. 
The OD has an 18 kton water volume monitored by 1885 outward-facing 8-inch PMTs to provide an active veto against charged particles originating from outside the detector. 
The ID contains 32 kton of water surrounded by 11,129 20-inch inward looking photomultiplier tubes (PMTs). 
To reduce counts from radioactivity from the PMTs and the mounting structure, a fiducial cut may be used to reject events $<$ 2 m from the ID walls, leaving a 22.5 kton fiducial volume. This ID fiducial volume is used in SN burst neutrino detection.

A neutrino interaction in the water volume produces a relativistic charged particle that creates a burst of Cherenkov photons which is detected by individual PMTs. The timing, $T$, and charge, $Q$, information from hit PMTs are continuously read out and associated with individual events based on software triggers. The $T$ and $Q$ from all hits in an event are used to reconstruct the event vertex, $\vec{x}$, as well as the energy, $E$, and direction, $\hat{d}$, of the outgoing particle. 

From 1996 to 2020, the SK~detector volume used ultra-pure water, but starting in the summer of 2020, gadolinium (Gd) was introduced into the detector by staged dissolution of the compound, Gd$_2$(SO$_4$)$_3$ $\cdot$ 8 H$_2$O, inaugurating the ``SK-Gd'' experimental era. 
This was done to increase the efficiency of detecting free neutrons produced by neutrino interactions, such as inverse beta decay (IBD) ($\bar{\nu}_e + p \rightarrow e^+ + n$)~\citep{Beacom:2004, MARTI2020163549}. 
In pure water, effectively all free neutrons are captured by hydrogen nuclei with an average capture time of $\sim$ 200 $\mu$s. 
Capture is followed by the emission of a 2.2 MeV $\gamma$ ($n + p \rightarrow \gamma$) only produces on average $\sim$ 7 Cherenkov photons.
Methods using the 2.2 MeV $\gamma$ yielded neutron-capture detection efficiencies of 17\%--26\%~\citep{ZHANG201541, Abe_2022}.
In SK-Gd, the neutrons have a high probability of capture by Gd due to the large neutron-capture cross sections of $^{157}$Gd and $^{155}$Gd. 
Capture is followed by a cascade of $\gamma s$ with a total energy of $\sim$8 MeV. This is sufficient to reconstruct the event energy and the neutron-capture vertex. In the case of IBD, the relativistic positron is seen as a prompt event and the neutron may be seen as a delayed event with reconstructed energy $<$10 MeV. 
Following individual event reconstruction, an IBD reaction may be identified by finding a coincidence between a prompt and a delayed event that are close in distance and time. If a delayed event from Gd $\gamma s$ cannot be matched with a prompt event, it may be misidentified as a prompt event. It is then considered a background (BGD) event.

The fraction of accurately tagged IBD events depends on the neutron-capture detection efficiency, the efficiency of coincidence identification and the background event rate. 
For the first phase of SKGd (SK-VI), a Gd mass concentration of 0.011\% was achieved, resulting in 50\% detection efficiency of neutron-capture on Gd with a mean capture time of $115 \pm 1$ $\mu$s~\citep{ABE2022166248}. 
For the second phase (starting with SK-VII), the Gd concentration was increased to 0.033\% resulting in a neutron-capture 
detection efficiency of $\sim$75\% with a mean capture time of $61.8 \pm 0.1$ $\mu$s~\citep{ABE2024169480}.

\subsection{Supernova Neutrino Detection at SK} \label{subsec:superksnneutrinos}
The detection of SN burst neutrinos is a major physics goal for the SK experiment with a long-running real-time SN neutrino burst monitor, \snww \citep{Abe:2016waf}. This system has recently been upgraded with several improvements to incorporate IBD tagging information and improve the speed of event reconstruction~\citep{Kashiwagi_2024}. In addition, changes in SN alert procedures and protocols have led to the implementation of a new, low-latency automated GCN SN burst notification system.

\begin{figure}[ht!]
\centering
\includegraphics[scale=0.4]{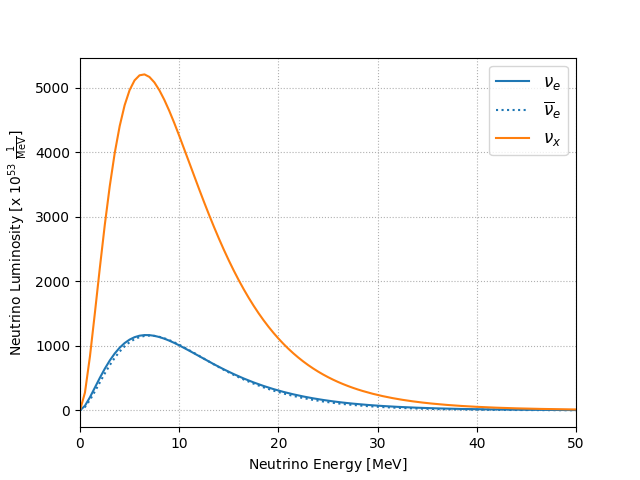}
\caption{The time-integrated neutrino energy spectrum for different flavors from the NK1 model with MSW oscillations  assuming normal mass ordering (see Section \ref{subsec:mcdatagen} for details). $\nu_x$ is the combined neutrino luminosity from $\nu_{\mu} + \bar{\nu}_{\mu} + \nu_{\tau} + \bar{\nu}_{\tau}$.
\label{fig:lum_nmo}}
\end{figure}

SN neutrinos have energies in the range of 0--80 MeV, with most below 50 MeV. At these energies, neutrinos
interact in the SK water volume by IBD, neutrino-electron elastic scatter (ES) 
($\nu + e^- \rightarrow \nu + e^-$),
charged-current (CC) reactions with $^{16}$O, (O16CC), 
($\nu_e + {\rm ^{16}O} \rightarrow e^- + X$ and 
$\bar{\nu}_e + {\rm ^{16}O} \rightarrow e^+ + X$)
and neutral current (NC) reactions with $^{16}$O, 
(O16NC), ($\nu + {\rm ^{16}O} \rightarrow \nu + {\rm ^{16}O^*}$). 
Figure~\ref{fig:lum_nmo} shows the energy distributions for different neutrino flavors calculated using the ``NK1'' SN model (see Section \ref{subsec:mcdatagen}) 
with MSW oscillations~\citep{Mikheyev:1985zog,Wolfenstein:1977ue} assuming normal mass ordering (NMO). For all flavors, the luminosities peak below neutrino energies of 10 MeV.

\begin{figure}[ht!]
\centering
\includegraphics[scale=0.5]{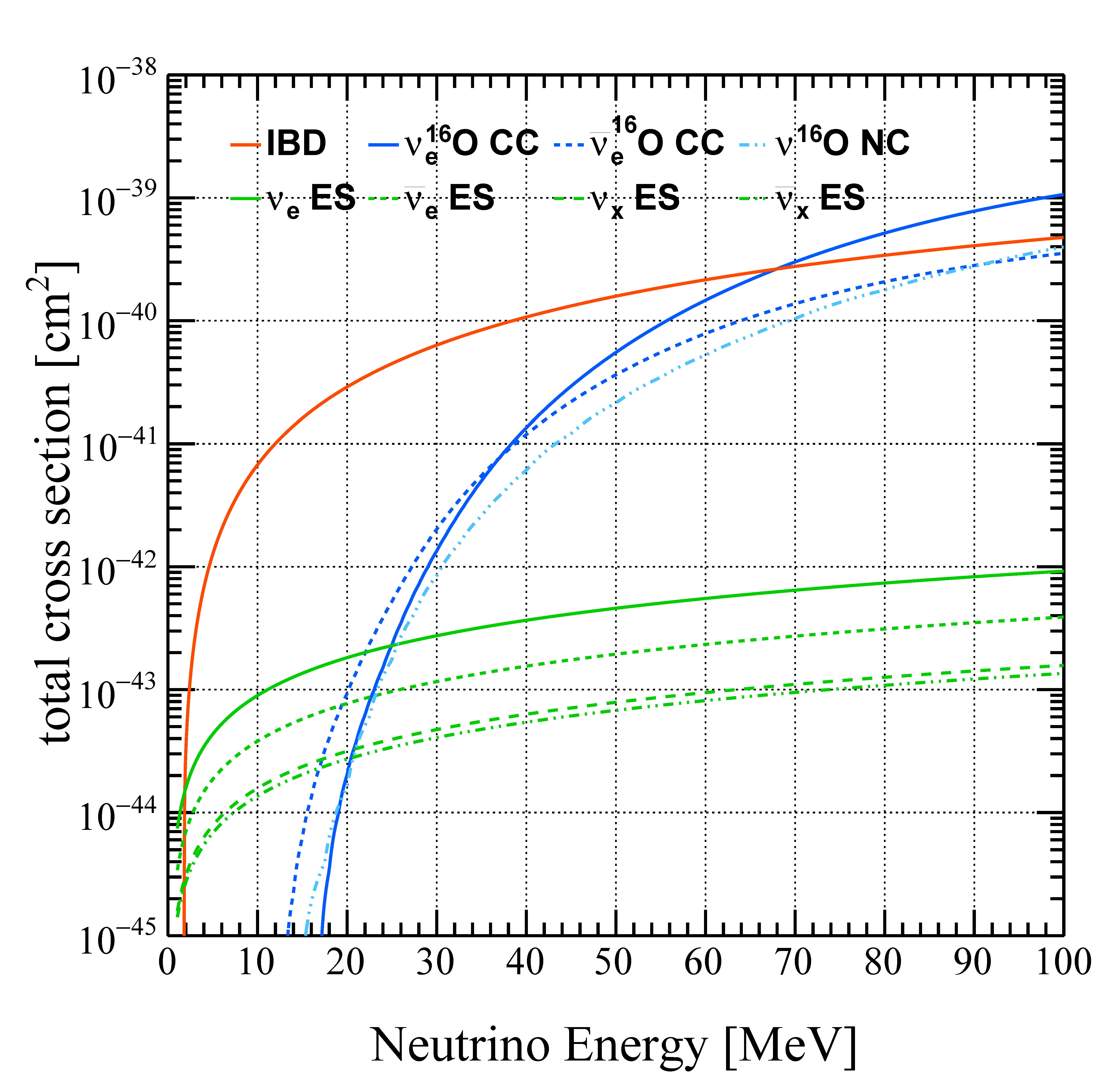}
\caption{The cross sections for the relevant neutrino reaction channels in water, from~\cite{Kashiwagi_2024}. 
\label{fig:xsection}}
\end{figure}

The number and energy distribution of observed events per reaction channel from an SN burst depends on the energy and neutrino flavor-dependent neutrino fluence, 
the reaction cross sections and the detector response.
The cross sections for the different reaction channels are shown in Figure~\ref{fig:xsection}. 
IBD has the largest cross section for $E_{\nu}$ $<$ 70 MeV with the cross sections for the various ES reactions significantly lower. The O16CC and O16NC reactions have high thresholds but the cross sections increase rapidly with energy. 

\begin{figure}[ht!]
\centering
\includegraphics[scale=0.6]{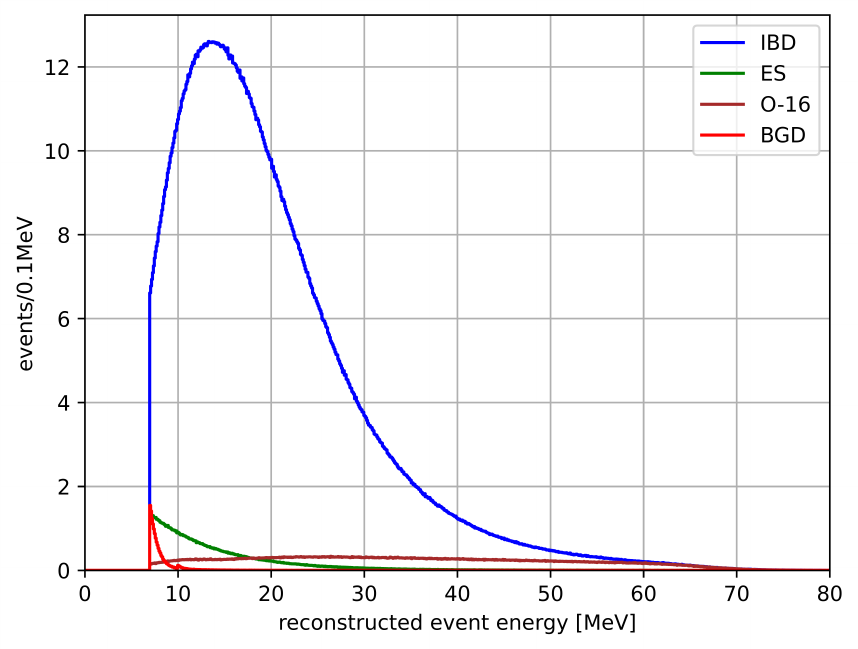}
\caption{The energy spectrum of reconstructed events in SK calculated using the NK1 model with NMO oscillations. The spectrum is subject to a 7 MeV energy threshold. BGD includes misidentified radioactive noise and spallation, but dominated by misidentified delayed events following Gd neutron-capture.}  
\label{fig:SKSNEventSpectra}
\end{figure}

Figure~\ref{fig:SKSNEventSpectra} shows the predicted energy distribution of observed events in SK for the different reaction channels calculated using the neutrino luminosities and energy distributions shown in Figure~\ref{fig:lum_nmo}. The observed energy spectra for IBD and O16CC events the track the neutrino energies, but the observed energies for electron ES events follow a recoil spectrum. 
IBD events dominate the spectrum for energies up to $\sim$70 MeV. The summed O16CC events are significantly lower, but also range from 0 to $\sim$70 MeV. The summed ES event distribution is highest at low energies, but negligible above $\sim$30 MeV. In this figure, BGD refers to misidentified events from background or following Gd neutron-capture. These have energies $<$10 MeV which partially overlaps with ES distribution.  

Table~\ref{tab:sknumevents2} shows the predicted average number of observed events per burst in SK for individual reaction channels calculated using two SN models and two types of neutrino oscillations for a SN distance of 10 kpc. These numbers are subject to a 7 MeV energy threshold. The event numbers for individual ES reaction channels (``ES1'', ``ES2'', ``ES3'' and ``ES4'') are listed separately. In all scenarios, the IBD events dominate. The total O16CC events is higher than the total ES, but the differences are more model-dependent. No O16NC events are listed because their energies are $<$ 7 MeV~\citep{10.1093/ptep/pty134}. Calculations of the event numbers in SK from other SN models are given in~\cite{Kashiwagi_2024}. 

\begin{table}[htbp]
\caption{The average SN burst events per reaction channel in SK for a SN at 10 kpc for the ``NK1'' and ``NK2'' SN models and ``NMO'' and ``IMO'' oscillations schemes from SKG4 simulations (see Section \ref{subsec:mcdatagen} for details). This includes only events flagged with sufficient reconstruction ``goodness'' and  $E > 7$ MeV.}
\begin{center}
\begin{tabular}{clccccc}
\hline \hline 
Label & Reaction & \multicolumn{2}{c} {NK1} & & \multicolumn{2}{c}{NK2}\\
& & NMO & IMO & &  NMO & IMO \\
\hline
IBD & $\bar{\nu}_e + p \rightarrow e^+ + n$ & 2434 & 2920 & & 2252 & 2652 \\
{\esnue} & $\nu_e + e^- \rightarrow \nu_e + e^-$      & 57 & 53 & & 54 & 54\\
{\esnuebar} & $\bar{\nu}_e + e^- \rightarrow \bar{\nu}_e + e^-$ & 11 & 13 & & 11 & 13\\
{\esnux} & $\nu_x + e^- \rightarrow \nu_x + e^-$ & 16 & 17 & & 17 & 17\\
{\esnuxbar} & $\bar{\nu}_x + e^- \rightarrow \bar{\nu}_x + e^-$ & 14 & 13 & & 13 & 14\\
{\oxccnue} & $\nu_e + {\rm ^{16}O} \rightarrow e^- + X$ & 107 & 83 & & 72 & 68\\
{\oxccnuebar} & $\bar{\nu}_e + {\rm ^{16}O} \rightarrow e^+ + X$ & 41 & 69 & & 41 & 64\\
\hline
& total & 2680 & 3168 & & 2460 & 2882\\
\hline \hline
\end{tabular}
\label{tab:sknumevents2}
\end{center}
\end{table}

The event numbers and energy distributions for some reaction channels are not independent. 
The IBD, {\esnuebar} and {\oxccnuebar} events are correlated because they depend on the $\bar{\nu}_e$ spectrum and fluence. 
In addition, {\esnue} and {\oxccnue} events are also correlated because both depend on the $\nu_e$ spectrum and fluence. 

The directional signal in an observed neutrino burst, used for SN direction reconstruction, depends on the magnitude of the asymmetry in the angular distribution of the observed events in the direction of the SN neutrino wavefront, \dnusn. The events in each reaction channel have a different angular distribution which may also vary with observed event energy. 
For IBD reactions involving $E_{\nu} < $15 MeV, the angular distribution of positrons is slightly asymmetric opposite to \dnusn. At higher energies, the bias reverses direction in the angular distribution reverses and its magnitude increases with energy but remains slight~\citep{PhysRevD.60.053003}. 
For SN neutrinos, the overall asymmetry depends on the detector energy threshold~\citep{STRUMIA200342, Ankowski_2019}. 
For O16CC events, the angular distribution of the outgoing particles is always biased opposite to the neutrino direction, but the magnitude of the bias depends on the energy threshold. 
For ES events, the outgoing electron directions are strongly forward scattered around \dnusn,
therefore the ES events provide the signal for SN direction reconstruction~\citep{Tomas:2003xn}.

\begin{figure}[ht!]
\centering
\includegraphics[scale=0.6]{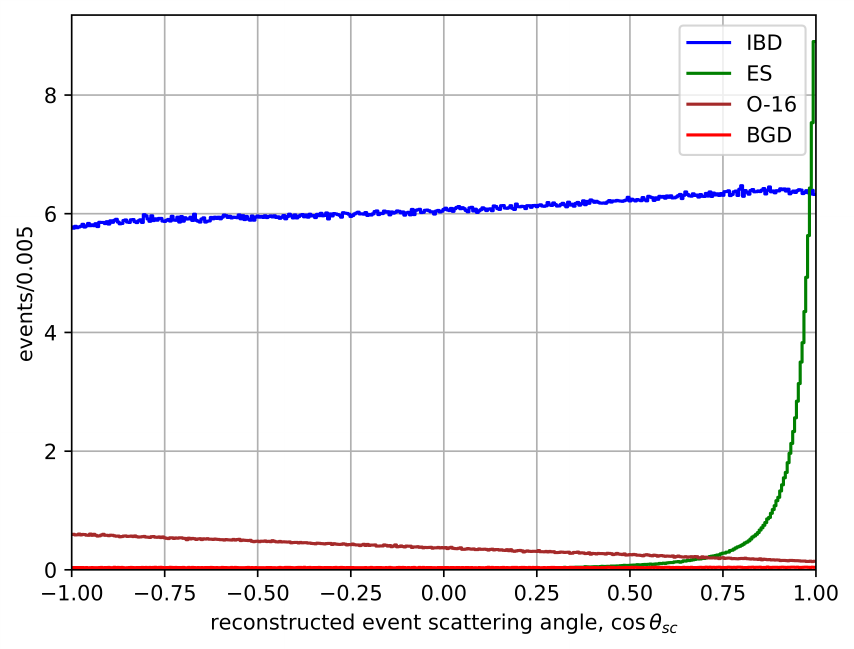}
\caption{The angular distribution of reconstructed particle directions for different neutrino reaction channels in SK for the NK1 SN model with NMO oscillations (see Section \ref{subsec:mcdatagen}). The IBD and O16CC events are nearly isotropic. The ES event directions are strongly forward biased. The artifactual BGD events show no directional bias.}
\label{fig:SKSNEventAngDist}
\end{figure}

Because SK is sensitive to all these reaction channels, the 3-d angular distribution of the burst events detected by SK has a directional signal from which the SN direction may be reconstructed. The angular distribution of the ES events (ES1 + ES2 + ES3 + ES4) provide the directional signal in the burst, while non-ES events (IBD + O16CC1 + O16CC2) are the directional background. 

When SNWATCH detects a burst of SN neutrinos, 
the reconstructed directions and energies of the burst events are used to reconstruct the SN direction \citep{Abe:2016waf}. SNWATCH used a maximum-likelihood fitter with an extended likelihood function with the SN direction angles and some reaction channel event numbers as fit parameters. Although ``ML-Fitter(2016)'' was able to calculate the SN direction, the computation times were long, increasing the latency of the SK SN alert system. Therefore, new methods were investigated to replace or improve the direction ``fitter'' in order to reduce the computation time. In addition, the new IBD tagging information was incorporated to improve the reconstruction accuracy by reducing the background from IBD events. These efforts were very successful, as detailed in the following sections. 

\section{HEALPix-based Method for SN Direction Reconstruction} \label{sec:healpixfitter}
\subsection{Overview} \label{subsec:hpoverview}
As described above, the outgoing particles from different reaction channels relevant to SN neutrinos have different angular distributions. Non-ES events have a roughly isotropic angular distribution but the ES events are strongly forward-scattered. In 3-d, the ES events are clustered around \dnusn, 
providing the directional signal and the non-ES events create a background~\citep{Beacom:1998fj, Tomas:2003xn}. 

A novel SN direction reconstruction method was developed that uses a HEALPix sphere as a data structure to encode and analyze the 3-d angular distribution of the SN burst events. This method, the ``HP-Fitter'', is extremely fast and with accuracy comparable to the standard SK maximum likelihood based fitter, the "ML-Fitter".

\subsection{HEALPix} \label{subsec:healpix}
HEALPix (Hierarchical Equal Area isoLatitude PIXelisation) is a scheme for encoding 3-d directional information using pixels that tile the 2-d surface of a sphere~\citep{2005ApJ...622..759G}. As such, it is widely used as a data structure for visualization and analysis of spherical directional data in geophysics, astrophysics, and cosmology, most notably for the cosmic microwave background. 

In the HEALPix architecture, equal area pixels are arranged in rings around the polar axis of sphere at constant latitudes. The pixels have equally spaced centers, except near the poles. The base HEALPix sphere has 12 pixels; three rings at different latitudes, each with four pixels per ring. The pixel shapes are roughly parallelograms with boundaries that do not follow the geodesic lines. The number of pixels is increased by subdividing each pixel into four equal-area pixels with the same properties. This gives HEALPix the ability to equivalently encode multiple data sets with different resolutions. 

The number of pixels on a \hp sphere depends on the number of times the base pixels are subdivided, $K$, and is controlled by the parameter NSIDE, where NSIDE$=2^{K}$. The impact of different values of NSIDE on pixel properties are
listed in Table~\ref{tab:healpixparams}. 

\begin{table}[htbp]
\caption{The HEALPix sphere properties for different values of NSIDE.}
\begin{center}
\begin{tabular}{cccc}
\hline
NSIDE & \# of Pixels & Angular Res & Pixel Area \\ 
& & (deg) & (deg$^2$) \\ \hline \hline
8 & 768 & 7.3 & 53.7 \\
16 & 3072 & 3.7 & 13.4 \\
32 & 12,288 & 1.8 & 3.4 \\
64 & 49,152 & 0.92 & 0.84 \\
128 & 196,608 & 0.45 & 0.21 \\
256 & 786,432 & 0.23 & 0.052 \\
\hline
\end{tabular}
\label{tab:healpixparams}
\end{center}
\end{table}

The use of \hp is supported by software libraries and packages that provide powerful tools, such as spherical harmonic analysis and spherical convolutional neural networks (CNNs) for machine learning~\citep{deepsphere_cosmo}. As well, Python wrappers, such as ``Healpy'', are available~\citep{Zonca2019}.

\subsection{Mapping SN Burst Events to a HEALPix Sphere} \label{subsec:evhealpix}
As described in Section~\ref{subsec:superksnneutrinos}, when an SN burst is detected in SK by SNWATCH, each event is reconstructed to calculate the interaction vertex and the energy and direction of the outgoing charged particle. In addition, information from Gd-neutron-capture is used to identify and tag IBD events. Since all SN burst neutrinos reaching the earth have the same direction, the 3-d angular distribution of the reconstructed event directions forms a ``scattering'' pattern around \dnusn. 

\begin{figure}[ht!]
\includegraphics[scale=0.40]{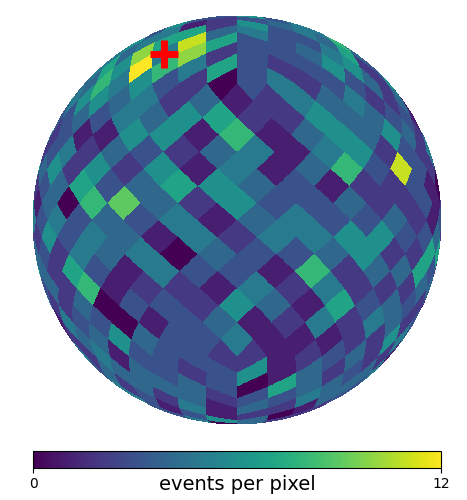}
\includegraphics[scale=0.35]{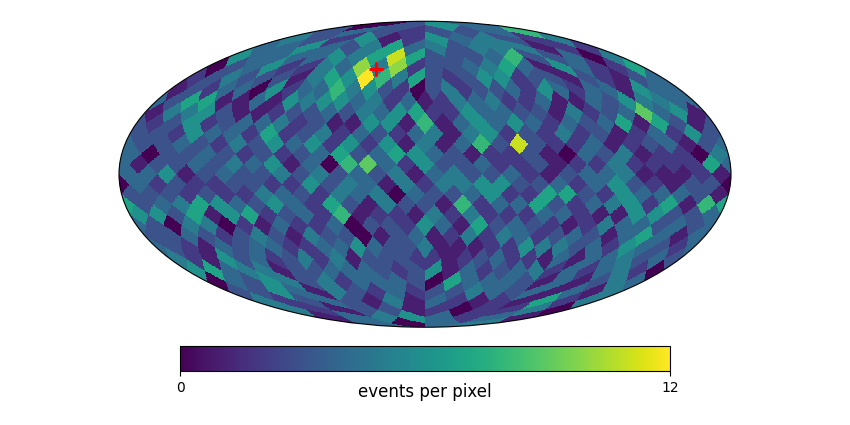}
\centering
\caption{A \hp event map with 768 pixels (NSIDE=8) loaded with $\sim$2700 burst events and visualized as an orthographic projection (left) and a Mollweide projection (right). The red cross shows the direction of the SN neutrino wavefront, \dnusn. The color bar shows the number of events per pixel.}
\label{fig:hp_sphere}
\end{figure}

The locations of the pixel centers on a \hp sphere correspond to 3-d direction vectors and the pixel area defines a solid angle on a unit sphere. 
Therefore, the set of burst events can be mapped onto a \hp sphere by setting the value of each pixel to the number of events that scatter into the solid angle subtended by that pixel. 
Other event information, such as reconstructed energy and IBD-flag status, may be used to make cuts on the events used to generate \hp map.
Figure~\ref{fig:hp_sphere} shows a \hp event map with 768 pixels (NSIDE=8) loaded with $\sim$2700 burst events visualized as an orthographic projection and Mollweide projection (sometimes called a "sky-map"). 

\begin{figure}[ht!]
\includegraphics[scale=0.4]{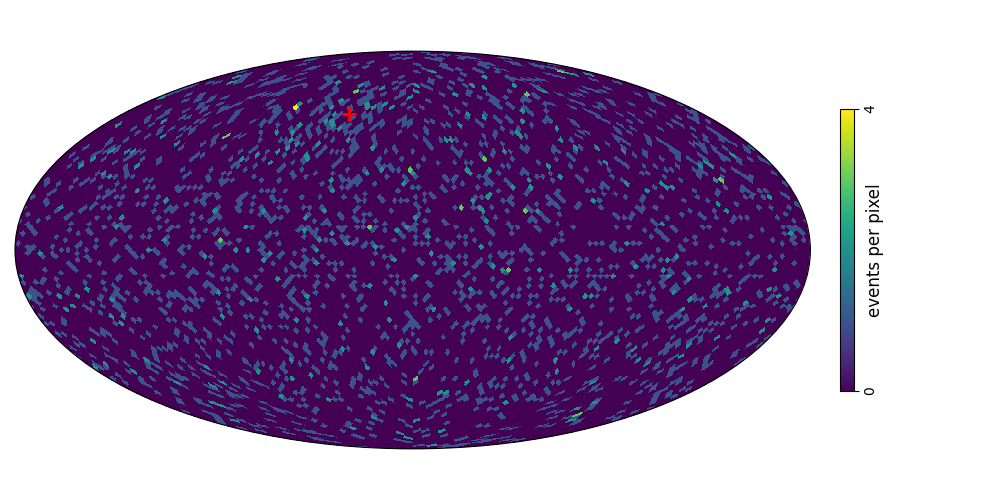}
\centering
\caption{The Mollweide projection of a \hp burst event map with 12,288 pixels (NSIDE=32) loaded with $\sim$2700 burst events. The red cross shows the direction of the SN neutrino wavefront, \dnusn. The color bar shows the events per pixel.
\label{fig:hpsphere3000events}}
\end{figure}

The burst event angular distribution is better visualized by using more pixels. Figure~\ref{fig:hpsphere3000events} shows $\sim$2700 burst events on a \hp map with 12,288 pixels (NSIDE=32). This map is sparsely populated with most pixels having no events. The distribution of events appears roughly uniform, with a slightly higher density around \dnusn.

\begin{figure}[ht!]
\includegraphics[scale=0.45]{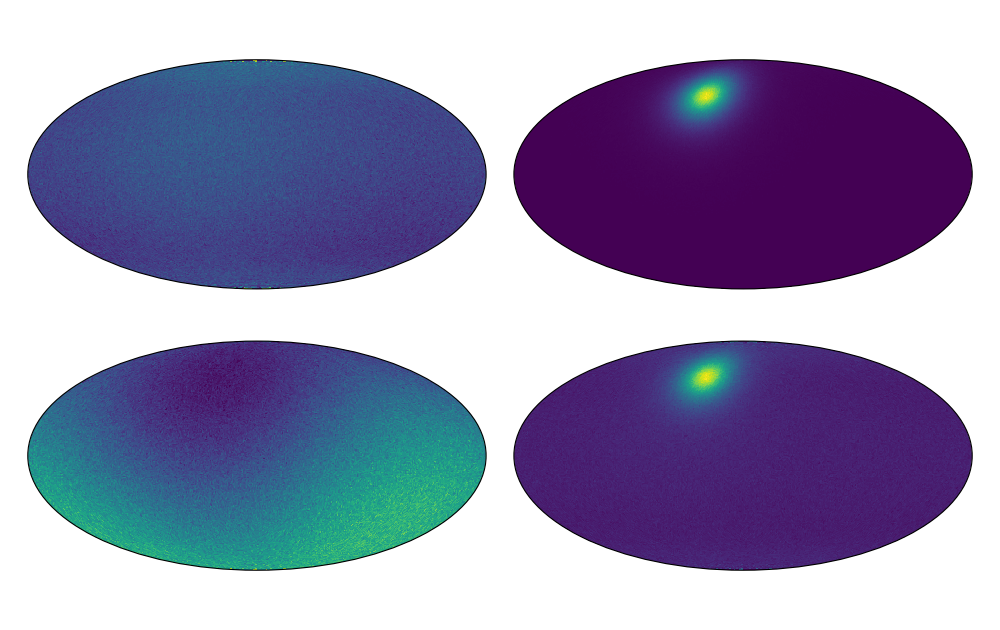}
\centering
\caption{The 3-d angular distributions of IBD (upper left), ES (upper right), O16CC (lower left) and IBD + ES + O16CC events (lower right). Different image scales were used to enhance the variations.}
\label{fig:evnt_types_on_hp}
\end{figure}

SN events from different reaction channels contribute to the \hp event map according to their 3-d angular distributions, as shown in 
Figure~\ref{fig:evnt_types_on_hp}. 
The IBD event distribution is nearly isotropic, with a slight bias in the direction of \dnusn. 
The O16CC events are also nearly isotropic, but with a slight bias opposite to \dnusn.
ES events are highly concentrated in the region around \dnusn.
The combined 3-d angular distribution from all burst events has a singular focal region of higher density, here designated the ``ES-peak''. This is the directional ``signal'' of an SN burst with a roughly isotropic ``background''. 

The 3-d angular distribution in a \hp event map may be characterized by the regional differences in the average events per pixel, here called the ``count density'' (C.D.). The absolute C.D. describes the sparseness of a region and determines the magnitude of the random variations in individual pixel counts, here called the ``noise''. 
The relative differences in regional C.D. determine the ``contrast'' of a given feature. 
In this context, the contrast between the ES-peak and the background depends on the difference in C.D. between the peak region and the background. The visibility of the ES-peak as depends on the magnitude of the noise in the peak and background regions relative to the contrast, here called the ``contrast-to-noise ratio'' (CNR). 
The \hp event map in Figure~\ref{fig:hpsphere3000events} shows a nearly uniform C.D. with only a hint of higher C.D. in the region around \dnusn\ but no distinct ES-peak. All regions of the map also have significant noise, due to the low C.D., that obscures the ES-peak. This map may be described as having a low CNR for the ES-peak.

\begin{figure}[ht!]
\includegraphics[scale=0.45]{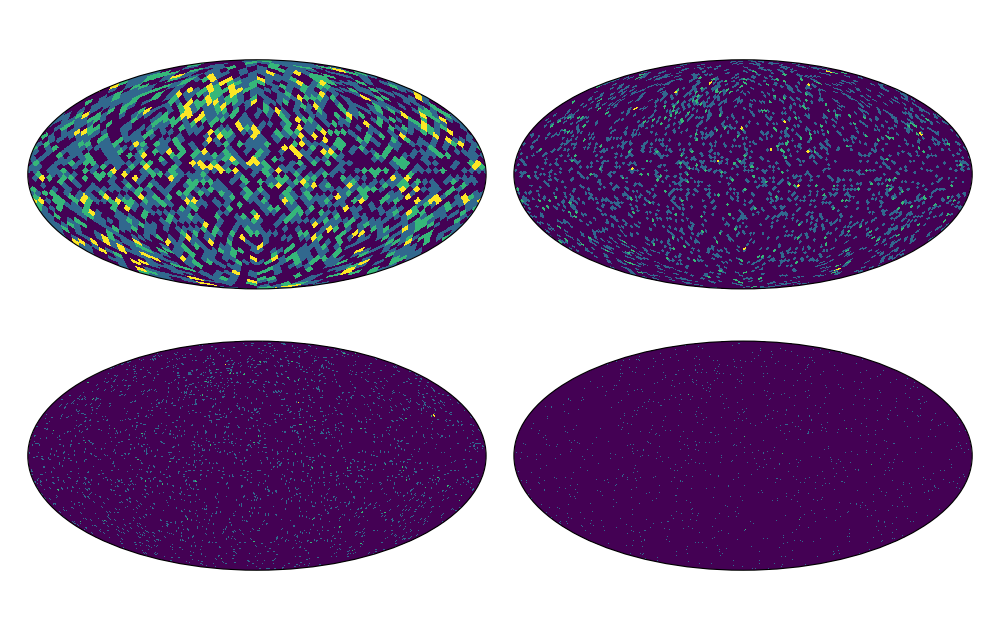}
\centering
\caption{Comparison of \hp maps for a burst with $\sim$2700 events with different values of NSIDE. The map with \ns=16 has 3072 pixels (upper left), \ns=32 has 12,288 pixels (upper right), \ns=64 has 49,152 pixels (lower left) and \ns=128 has 196,609 pixels. As NSIDE increases, the maps become increasingly sparse. 
\label{fig:hpsphere_comp_res}}
\end{figure}

Since the C.D. of a \hp map depends on the number of events and the number of pixels, the ES-peak CNR should vary with the choice of NSIDE. Lowering NSIDE reduces the number of pixels resulting in higher C.D., which should increase the CNR of the ES-peak.
Figure~\ref{fig:hpsphere_comp_res} shows \hp maps with the same number of burst events but different values of NSIDE. Maps with lower NSIDE are less sparse, but since the number of burst events is so low, there is no significant improvement in the CNR of the ES-peak. In addition, maps with lower NSIDE suffer from worse pixel angular resolution as noted in Table~\ref{tab:healpixparams}. This would ultimately limit the precision of SN directions found using the map.

\subsection{Extracting the SN Direction from HEALPix Event Maps} \label{subsec:dirhealpix}
Our initial strategy for SN localization was to train a graph spherical convolutional neural network (CNN), ``DeepSphere'' \citep{deepsphere_cosmo}, to relate the image patterns on a \hp event map to the SN direction. 
Although initial tests looked promising \citep{10.1145/3408877.3439692}, this method could not produce consistent or reliable results. 
This led us to search for ways to increase the CNR of the ES-peak in the \hp map. 
As described above, reducing NSIDE to increase the C.D., had negligible impact on ES-peak CNR due to the overall sparseness of the maps. 

\begin{figure}[ht!]
\includegraphics[scale=0.50]{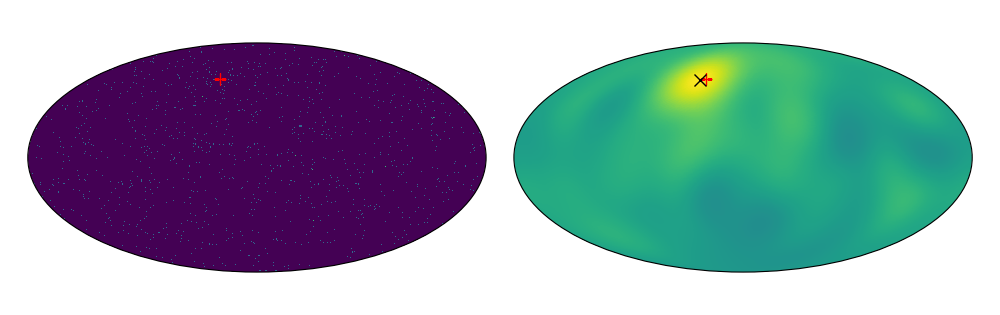}
\centering
\vspace{4pt}
\caption{A \hp map with NSIDE=128 loaded with $\sim$2700 burst events before (left) and after (right) Gaussian smoothing. The red ``+'' indicates the true \dnusn. A prominent ES-peak is revealed in the smoothed \hp map around \dnusn. The ES-peak centroid is located at the maximum value pixel (black ``$\times$'') and is used to determine \dnusnrecon.
\label{fig:comp_smhp}}
\end{figure}

The solution was found by applying Gaussian smoothing to the \hp event map. 
Smoothing is commonly used in medical imaging to increase the CNR for lesion detection (e.g., \cite{10.1007/978-981-16-9113-3_34}). 
Gaussian smoothing convolutes the sparse event distribution with a 2-d Gaussian kernel to effectively smear individual events over several pixels. 
The overlap of smeared events results in a smoothed map with pixel values that reflect the regional ``information density'' (I.D.), in events per steradian, 
and yields a better characterization of the 3-d angular distribution of the burst events. 
In imaging terms, smoothing reduces the high spatial frequency contributions to the image power spectrum (dominated by noise) to reveal the underlying low frequency features of the 3-d event angular distribution, most notably the ES-peak. 
The results are also less dependent on pixel parameters, such as pixel angular resolution, and therefore less dependent on the choice of NSIDE. 
The impact of smoothing on a sparse \hp event map is dramatic, as shown in Figure \ref{fig:comp_smhp}.

\begin{figure}[ht!]
\includegraphics[scale=0.45]{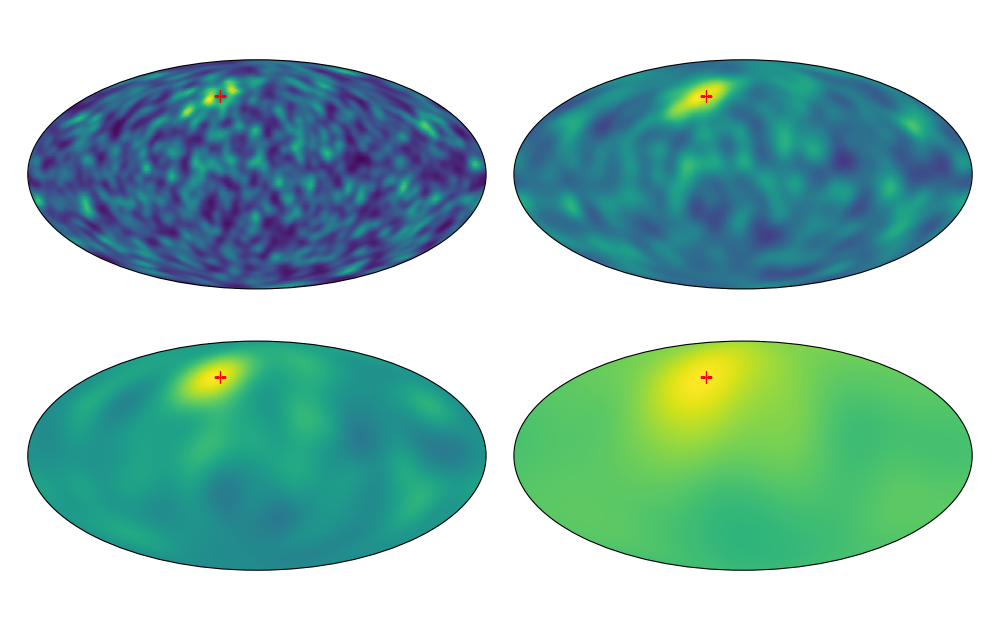}
\centering
\caption{Comparison of \hp maps with NSIDE=128 smoothed using SIGMA=0.05 (upper left), 0.10 (upper right), 0.20 (lower left) and 0.40 radians (lower right). The red ``+'' indicates the true \dnusn. Increasing SIGMA reduces the background fluctuations revealing the ES-peak. Higher values of SIGMA broaden the ES-peak, making the peak centroid less distinct and reducing the contrast.
\label{fig:hpsphere_comp_smooth}}
\end{figure}

The amount of smoothing depends on the width of the gaussian function (in radians), and is controlled in \hp by the SIGMA or the FWHM parameter. Figure~\ref{fig:hpsphere_comp_smooth} shows \hp maps smoothed with different values of SIGMA. Without adequate smoothing, the ES-peak may not be clearly differentiated from peaks created by background fluctuations. Increased smoothing makes the ES-peak more distinct, with a clearly defined centroid. Over-smoothing may reduce the ES-peak contrast and make its centroid less defined. Optimal smoothing should produce a solitary, narrow peak.

Since the ES-peak is formed from ES events with small scattering angles, the peak centroid should be nearly co-directional to \dnusn. In a smoothed \hp map, the centroid should be located close to the pixel with the highest amplitude. 
Therefore, the direction of the SN neutrino wavefront may be reconstructed by simply finding the direction corresponding to maximum pixel in the smoothed \hp map, \dnusnrecon$ = \hat{d}_{pixel,max}$. 
Finally, since the direction of the SN neutrino wavefront is opposite to the sky-map location of the SN (\dsntrue\ = $-$\dnusn), the reconstructed pointing direction for SN localization is found from \dsnrecon = $-$\dnusnrecon. 
This method of SN localization is designated the ``HP-Fitter''. Figure \ref{fig:comp_smhp} shows the raw and smoothed \hp maps with the "true" \dnusn\ and the   \dnusnrecon\ reconstructed by the HP-Fitter.

The precision of \dsnrecon\ determined by the HP-Fitter depends on the \hp map pixel angular resolution, which is controlled by NSIDE, as listed in Table~\ref{tab:healpixparams}. For example, if NSIDE=128 is used, the pixel angular resolution is 0.45\degree.
The accuracy of \dsnrecon\ for individual bursts depends fundamentally on the number and directions of the ES events that form the ES-peak, as well as the number and directions of the non-ES events in the background. 
Since these are subject to random variations, the overall accuracy of the HP-Fitter accuracy will depend on the absolute and relative numbers of ES and non-ES events in a burst. This implies that the accuracy will vary with SN distance and between different SN neutrino flux models. The accuracy will also depends on the choice of HP-Fitter parameters: NSIDE and SIGMA. These parameters must be optimized for different burst observables, such as the total number of burst events. HP-Fitter parameter optimization is described in Section \ref{subsec:results-hp-opt}.

\begin{figure}[ht!]
\includegraphics[scale=0.25]{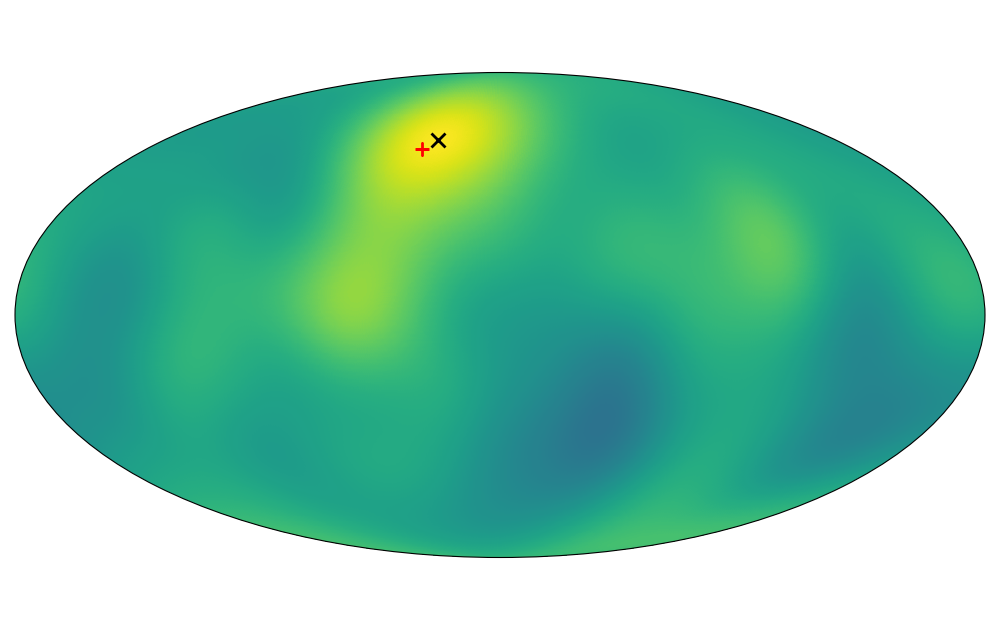}
\includegraphics[scale=0.25]{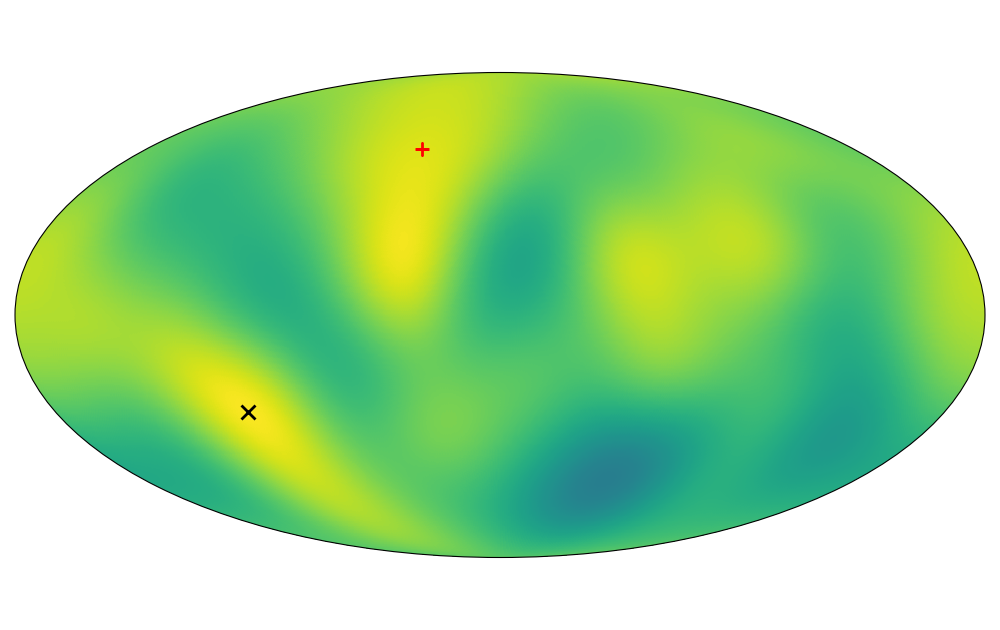}
\centering
\caption{Comparison of smoothed \hp burst event maps from two bursts at 20 kpc generated from the same SN flux model. The red ``+'' indicates the true  direction, \dnusn, and the black ``$\times$'' indicates the reconstructed direction, \dnusnrecon. One burst (left) shows a distinct ES-peak which yields an accurate \dnusnrecon, with an angular discrepancy of only 5.02\degree~from \dnusn. A second burst (right) does not produce a distinct ES-peak and \dnusnrecon\ was determined based on a random peak in the background. This results in a failed reconstruction, with an angular discrepancy of 98.7\degree.} 
\label{fig:hpfitter-fail}
\end{figure}

The HP-Fitter only produces a meaningful estimate of \dsntrue\ if the ES-peak has the highest amplitude on the smoothed \hp map. Other peaks appear on the map from random fluctuations in the background event densities. If a background peak has higher amplitude than the ES-peak, \dsnrecon\ is determined based on the random location of the background peak, resulting a ``failed'' direction reconstruction. This may occur if the \hp event map is not smoothed adequately or if the absolute and/or relative number of ES events in an individual burst is too small to produce a ES-peak of sufficient CNR. 
Figure~\ref{fig:hpfitter-fail} shows smoothed \hp event maps for two instances of a SN burst at 20 kpc simulated with the same SN neutrino flux model. The bursts differ only by random variations in the ES and non-ES event numbers, energies and directions. Both maps used the same NSIDE and SIGMA. The \hp map for one burst (left) shows a distinct ES-peak with high CNR and the fitter produces an accurate \dsnrecon. The \hp map for the other burst (right) does not have a distinct ES-peak with higher amplitude than the random background peaks. In this instance, the HP-Fitter fails and the output \dsnrecon\ is random, unrelated to \dsntrue.  

\begin{figure}[ht!]
\includegraphics[scale=0.50]{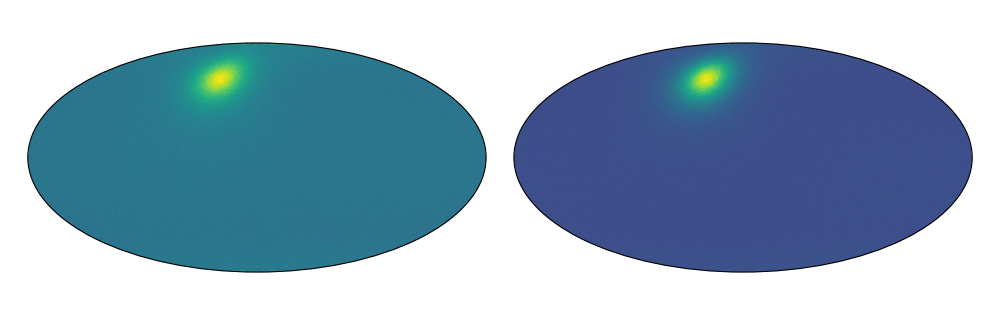}
\centering
\caption{Burst event angular distributions before (left) and after (right) the removal of IBD-tagged events showing the improvement in ES-peak contrast.}
\label{fig:ibd_flagged_sub_hp}
\end{figure}

Since the HP-Fitter's accuracy and likelihood of failure depend on the ES-peak CNR, reducing the number of non-ES events in the \hp map should yield improvements in both. 
With the advent of SK-Gd, a large fraction of IBD events are identified and tagged during event reconstruction. This allows events identified as IBD to be excluded from the \hp event map, which should increase the contrast of the ES-peak. Figure~\ref{fig:ibd_flagged_sub_hp} compares the burst event angular distributions before and after removing the IBD-tagged events, showing the improvement in ES-peak contrast. This cut was found to both improve accuracy and reduce failures. 
Since most SN burst events with $E_{recon}>30$ MeV are from IBD or O16CC reaction channels (see Figure~\ref{fig:SKSNEventAngDist}), various acceptance cuts based on event energy were also tested. However, there was found to not have a significant impact on HP-Fitter performance. 

\subsection{Implementation} \label{subsec:hp-impl}
The HP-Fitter was coded in Python and made extensive use of many Python packages and libraries,  
especially ``Healpy''~\citep{2005ApJ...622..759G, Zonca2019}. 
The code was not compiled and no GPU or hardware acceleration was used. 
The HP-Fitter was optimized by finding the values of NSIDE and SIGMA that optimize the angular resolution and minimize the failure rate, as described in Section \ref{subsec:results-hp-opt}.

\section{Improvements to the Maximum-Likelihood Fitter} \label{sec:mlfitterimprov}
\subsection{Introduction} \label{subsec:mlfitterintro}
The original SN direction reconstruction method employed by SNWATCH, ``ML-Fitter(2016)'', uses a maximum likelihood optimization with an extended likelihood function to estimate not only the SN direction, but also the number of events per reaction channel in coarse energy bins \citep{Abe:2016waf}. 
This method was computationally intensive, requiring several minutes to perform the fit. Furthermore, the computation time increases with the number of events in a burst, which could be problematic for a nearby SN. This computational delay was a significant bottleneck in the time required to produce the SK SN burst alert. 
In parallel with the development of the HP-Fitter, recent work was done to upgrade the ML-Fitter methodology and code which yielded significant improvements to its speed and accuracy. 

\subsection{ML-Fitter Code Improvements} \label{subsec:mlfittercodeimprovements}
The ML-Fitter uses an extended likelihood function with fourteen fit parameters: two SN direction angles and twelve estimates of the number of events per coarse energy bin for different reaction channels. The likelihood calculation uses a probability density function (PDF) for each reaction channel, $p_{r}(E_i; \theta_i)$, that relates the probability of the event scattering angle, $\theta_i$, to the observed event energy, $E_i$, where 
$\theta_i=\hat{d}_i \cdot$ \dnusn and $\hat{d}_i$ are the reconstructed event directions. The PDFs are constructed by fitting the MC simulated data to empirical equations. For IBD and O16CC events, the PDFs use simple empirical functions parameterized by $E_i$ and $\theta_i$. 
For ES events, separate PDFs are used for different energy slices to reduce the dependence on the SN flux model used to generate the simulated data. The PDF for each energy slice is parameterized by $\theta_i$ and the fit to the simulated data requires a complex, empirical formula. More details of the ML-Fitter method are given in Appendix \ref{sec:mlfitterdetails}. 

The first upgrade to ``ML-Fitter(2021)''~\citep{Kashiwagi_2024} incorporated IBD tagging information from SK-Gd. Initially, IBD tagged events were down-weighted in the likelihood calculation but later versions of the algorithm removed these events for better performance. This cut reduced the ``background'' in the angular distribution of the burst events used by the ML-Fitter and yielded improved directional accuracy.

Further work focused on reducing the computation time.
For ``ML-Fitter(2022)'', the code was refactored from C++ to Python. Since Python is basically a script language, it would be expected to be much slower than the compiled C++ used in earlier versions. 
However, Python has several powerful methods for code optimization that improve calculation speed. 
Extensive work was done to vectorize every part of the Python ML-Fitter code so that functions operated on the event data as arrays rather than individual events. 
This eliminated the widespread, nested looping used in the original C++ code, which was especially important for the likelihood function that would be called hundreds or thousands of times during likelihood optimization. 
Additional gains were made by pre-calculating variables outside of the likelihood function to reduce identical calculations during the optimization. 
The Python code also benefited by using existing packages and libraries that use optimized and/or compiled code to improve performance. For example, ``iMinuit''~\citep{iminuit} is an optimized Python wrapper for CERN's Minuit2 C++ library~\citep{James:1975dr} that was used for the likelihood maximization. Speed improvements were also made by using cached, compiled functions. 

ML-Fitter(2022) also benefited from using the SN direction calculated by the HP-Fitter for the starting values for the SN direction fitter parameters. Previously, a slow and less accurate ``grid-search'' was used, so the SN direction parameters began further from the optimal values. Using more accurate direction parameters reduced the number of iterations needed for optimization and the danger of the ML-Fitter stopping on a local maximum. These upgrades made ML-Fitter(2022) significantly faster and slightly more accurate than ML-Fitter(2021).

\section{Methods} \label{sec:methods}
\subsection{Supernova Burst Simulations} \label{subsec:mcdatagen}
Simulated SN burst events in SK were generated for use in the development, testing and optimization of the new and improved fitters. This requires several steps, described in more detail in~\cite{Kashiwagi_2024}. 
First, a SN model and an MSW neutrino oscillation scheme are chosen. These determine the time, energy and flavor-dependent luminosities of the SN neutrino emissions. The neutrino flux through the SK detection volume then depends on the choice of SN distance and sky location.
The neutrino fluxes and energy distributions are then used to calculate the characteristic energy spectra and angular distributions of events from each reaction channel based on the cross sections and the numbers and types of target atoms in the detector volume.
Next, the SK event generator, ``SKSNSim''~\citep{Nakanishi_2024}, is used to generate individual event instances from these distributions and create a simulated burst. Each event has a ``true'' value for the interaction vertex (location and time), and for the outgoing particle type, energy and direction. Each burst consists of a unique set of events, each with different characteristics. The event information for each burst is stored in a ``vector file''. Large numbers of burst vector files are generated, each with random variations in the event content.

The burst vector files are then used by ``SKG4'', the GEANT4-based Monte Carlo (MC) package for simulating particle interactions in the SK detector~\citep{Harada_2020}. 
For each event in the burst, SKG4 simulates detailed particle propagation, scatter and Cherenkov light production in the water volume, as well as the propagation, absorption, scatter and detection of the Cherenkov photons. These photons are tracked to simulate PMT hits and generate signal charge, $Q$, and timing, $T$, information for each PMT hit.
Realistic backgrounds are included by injecting dark noise and background PMT hits from random, wide trigger detector data into the sample of simulated hits. 
Since rejected detector event data is not stored, the background rate injected in the MC can only be estimated.
The injected noise PMT hits produce about 30\,000 events passing trigger conditions 
in a 20 sec burst, before applying event selection.
After selection, the background rate is estimated to be $\sim$6.7 events in a 20 sec burst, or $0.23 \pm 0.01\%$ of the selected sample from a burst at 10 kpc. 

The simulated PMT signal information ($Q$ and $T$) from SKG4 for each event is then used by the SNWATCH-modified version of the SK low energy event reconstruction program to calculate the vertex, and the direction and energy of the outgoing particle(s) for that event. With SK-Gd, the event information, together with the reconstruction quality parameters, is used to identify and tag correlated prompt and delayed events which are the signature of IBD with Gd neutron-capture. Delayed events from neutron-capture that are accidentally selected as prompt contribute to the background. For a SN burst at 10 kpc, these can add $\sim$32 background events per burst, or $1.15 \pm 0.02\%$ of the selected sample. This number scales with the number of IBD interactions.

For this work, the MC simulations used the SKVII detector configuration with 0.033\% Gd loading. The version of the SNWATCH analysis code had a IBD tagging fraction of 
$49.7 \pm 0.04\%$ with our datasets based on the same analysis as in \cite{Kashiwagi_2024}. The simulated events were subject to a fiducial cut that rejects events $\leq$2 m from the walls. An energy threshold of 7.0 MeV was used and only events with high reconstruction goodness criteria were used. 

Simulated data sets were generated using the Nakazato model~\citep{Nakazato_2013}. Two sets of parameters were available, ``NK1'' with a progenitor mass of 20\(M_\odot\), a shock revival time of 200 ms and 0.02 metallicity and ``NK2'' with a progenitor mass of 13\(M_\odot\), shock revival time of 100 ms and 0.004 metallicity. 
For the initial development, optimization and testing of the fitters only NK1 was used.
The effects of adiabatic MSW oscillations~\citep{PhysRevD.62.033007} assuming normal mass ordering ``NMO'' or inverted mass ordering ``IMO'' were also modeled.
Analysis of the simulated burst event detection in \snww for a wider range of SN models is given in~\cite{Kashiwagi_2024}. 
A single SN sky direction was used in the event generation, but, during testing, the reconstructed event directions were rotated to simulate other SN directions.
The SN burst events were generated for fixed SN distances. At least 3000 simulated bursts were generated for each distance for each SN flux model (SN model + mass ordering).

In order to simulate bursts with different parameters (distance, IBD tagging fraction, reaction channel content), all the burst events were combined into a single pool. 
Then, events could be randomly drawn from the pool based on the desired criteria for the resampled bursts. When simulating different SN distances or event numbers, the correct statistics were preserved by having the required number of events vary according to a Poisson distribution. 
The effects of high event rates on detector event processing were not modeled. 

\subsection{Angular Resolution} \label{subsec:fitterangres}
The accuracy of the SN direction reconstructed by a fitter is characterized by the angular discrepancy, \thsn
\footnote{$\Delta\theta_{SN}$ is used elsewhere in the literature for angular discrepancy.}, between the true SN direction, \dsntrue, and the reconstructed direction, \dsnrecon, where: 

\begin{equation}
\cos{\theta_{SN}} = \hat{d}_{SN} \cdot \hat{d}_{SN}^{recon}
\label{eq:def-cos-th}
\end{equation}

For a single burst, $\theta_{SN}$ depends on the numbers, directions and energies of the ES and non-ES events in the burst, as well as the fitter performance. 
However, different instances of a SN burst from the same neutrino flux will have random variations in the observed numbers and directions and energies of the ES and non-ES events in the burst. 
These will cause random variations in \dsnrecon, and therefore $\theta_{SN}$, between different burst instances. 
Therefore, the fitter performance can be characterized by the {\thsn} distribution from a large ensemble of simulated burst instances.
Each variation in the SN flux model, SN distance or in the fitter parameters will produce a different characteristic {\thsn} distribution. 

\begin{figure}[ht!]
\includegraphics[scale=0.5]{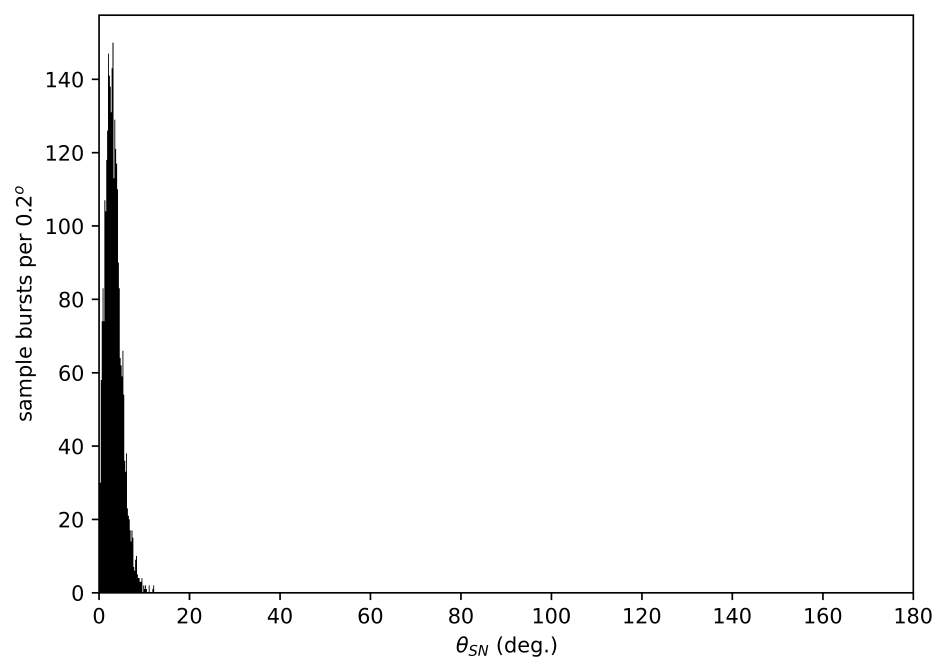}
\includegraphics[scale=0.5]{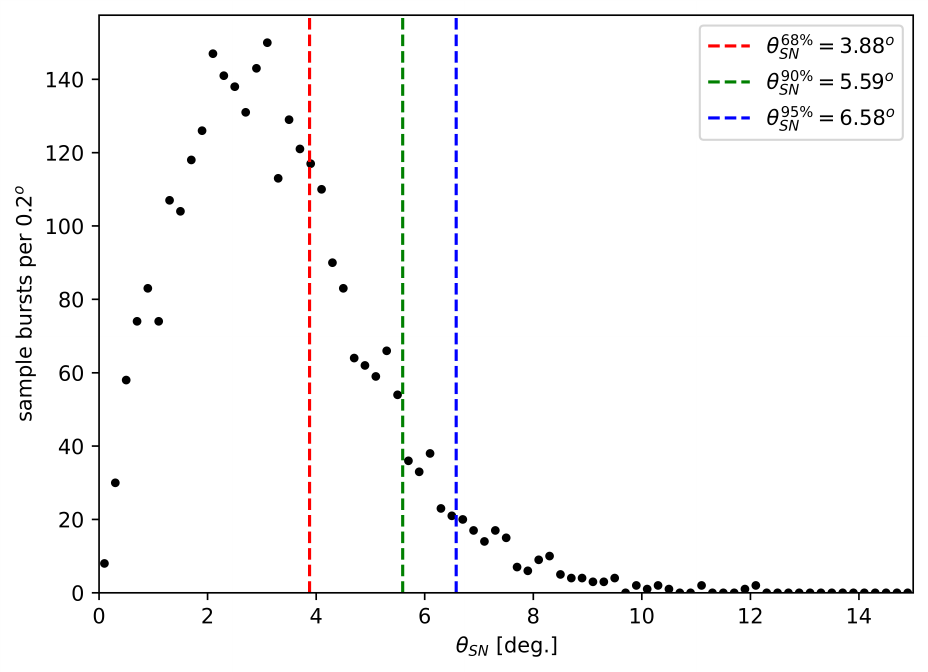}
\centering
\caption{The left figure shows the full {\thsn} distribution from the HP-Fitter for 3000 simulated bursts at 10 kpc. The right figure shows same distribution 
for $\theta_{SN}$ = 0\degree--15\degree\ with the extracted percentile angular limits.}
\centering
\label{fig:theta_sn_anal}
\end{figure}

To quantify the fitter accuracy (or ``angular resolution'') for a given set of parameters, the fitter is first used to reconstruct \dsnrecon\ for a large set 
of simulated bursts ($\geq$ 3000) generated using the same SN flux model and distance. For each reconstructed direction, \thsn\ is calculated and the characteristic {\thsn} distribution is the histogram created from the set of bursts. 
This can be analyzed to extract measures of the angular resolution, such as the average angular discrepancy, \avgdtheta, 
and the angular limit for a given percentile of the bursts, such as \dtheta{68}, \dtheta{90} and \dtheta{95}. 
These angles may be interpreted as the radii of error circles on a sky map that enclose the given fraction of the reconstructed SN directions from the fitter. They may also be interpreted as the probability of finding the true SN direction at a given angular distance from \dsnrecon. 
Since the {\thsn} distributions include the reconstructed SN directions from all bursts, \avgdtheta, \dtheta{68}, \dtheta{90} and \dtheta{95} may be skewed by failed reconstructions which produce random SN directions.
Figure~\ref{fig:theta_sn_anal} shows the full \thsn\ distribution (left) and the essential \thsn\ distribution for $\theta_{SN}$=0\degree--15\degree\ (right). Also shown are the angular limits for different percentiles.  

\subsection{Estimation of Reconstruction Failure Rate} \label{subsec:fitterfailurerate}
\begin{figure}[ht!]
\includegraphics[scale=0.5]{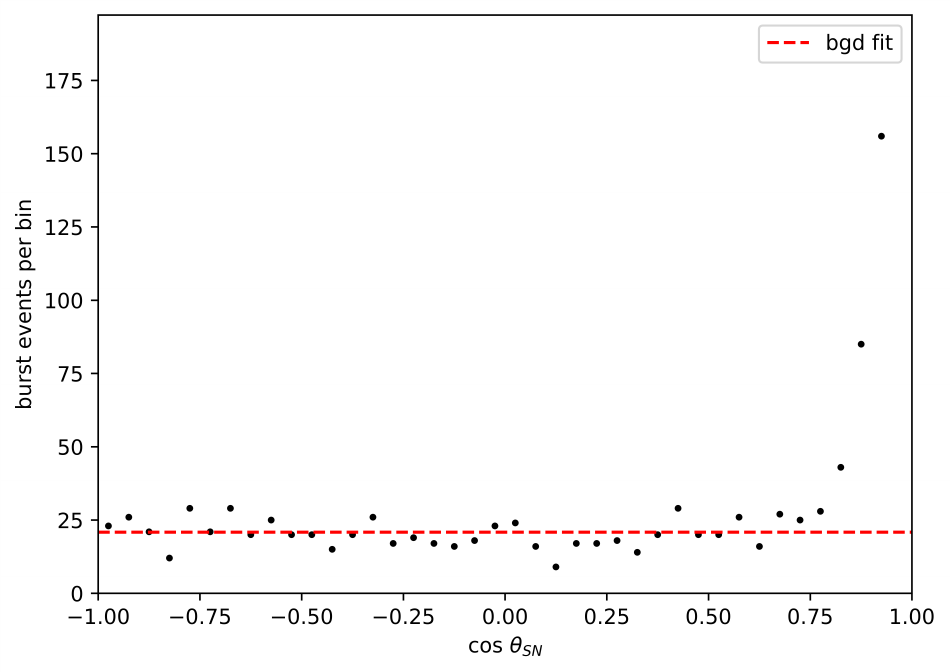}
\includegraphics[scale=0.5]{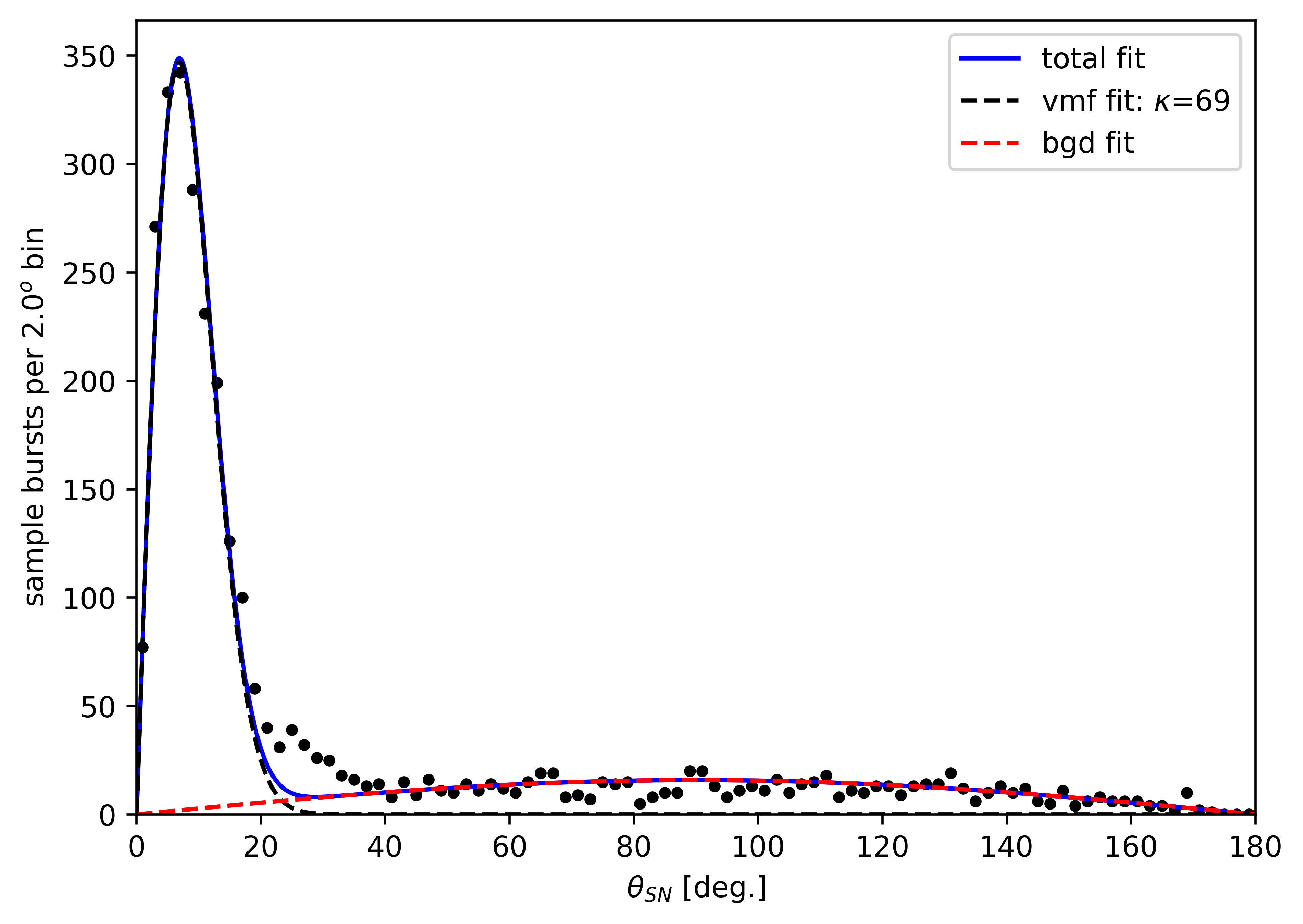}
\centering
\caption{The angular discrepancies of reconstructed SN directions from the HP-Fitter from simulated bursts at 30 kpc. Bursts at this distance have few ES events and therefore a higher failure rate. In the $\cos\theta_{SN}$ distribution (left), the random reconstructed SN directions from failed reconstructions produce a constant background. In the {\thsn} distribution (right), the failed reconstructions produce a background with the form of a half-sine wave. Also shown is a fit of the {\thsn} distribution to a von Mises-Fisher function and a half-sine wave background.}
\centering
\label{fig:vmf_fit_bgd}
\end{figure}

When reconstructing the SN direction from bursts with poor directional information (low absolute and/or relative number of ES events), the fitter can fail resulting in a random \dsnrecon\. 
For the HP-Fitter, this occurs when the smoothed ES-peak has a lower amplitude than a peak from random fluctuations in the non-ES background, as shown in Figure~\ref{fig:hpfitter-fail}. 
For the ML-Fitter, the direction reconstruction fails when the statistical fluctuations in the burst event content create a random global maximum (or local maximum close to the starting fitter parameters) in the likelihood function. 
This is more likely to occur when the ML-Fitter uses a random direction from a failed HP-Fitter reconstruction for the initial {\dsnrecon} fitter parameters.
Random $\hat{d}_{SN, recon}$ from failed reconstructions create a constant background in the {\cthsn} distribution and a half-sine wave background in the {\thsn} distribution, as shown in Figure~\ref{fig:vmf_fit_bgd}. 
% This {\thsn} distribution shows a large deviation from the von Mises-Fisher function from 20\degree--40\degree which is likely caused by the round 20-40 degrees understood?}

The fitter failure rate, $R_{fail}$, can be estimated from the constant background in the {\cthsn} distribution.
Since the {\cthsn} distribution is sharply forward biased, it may be assumed that all burst samples in the range {\cthsn} = $-1.00$ to 0.00 are from failed reconstructions. 
Therefore, the total number of failed reconstructions, $N_{fail}$, may be estimated by counting the number of burst samples in the range {\cthsn} = $-1.00$ to 0.00 and multiplying by two.

The fitter failure rate is then calculated from $N_{fail}$ and the total number of burst samples, $N_{total}$:

\begin{equation}
R_{fail} = \frac{N_{fail}}{N_{total}}
\label{eq:fract_fail}
\end{equation}

\section{Results} \label{sec:results}
\subsection{HP-Fitter Parameter Optimization} \label{subsec:results-hp-opt}
Since the HP-Fitter performance depends on the choice of NSIDE and SIGMA, tests were performed to determine the values of these parameters that optimize the angular resolution over a range of SN distances which varies the number of events per burst. 

A set of 3000 burst samples was generated for SN distances from 2--30 kpc by randomly selecting from the pool of simulated events. The average events per burst was scaled to account for the differences in distance.
For each burst sample, the HP-Fitter was used to reconstruct the SN direction using three values of NSIDE (64, 128, 256) and values of SIGMA from 0.10--0.50 rad. The SN directions were randomized to prevent biases based on the relationship between \dsntrue\ and the small sampling biases in the HEALPix pixel structure. 
This yielded a characteristic \thsn\ distribution for each (NSIDE, SIGMA) pair for each distance, from which \avgdtheta\ and \dtheta{68} were calculated. 
These specific angular resolution measures were used because they are least sensitive to failed reconstructions. 

\begin{figure}[ht!]
\includegraphics[scale=0.35]{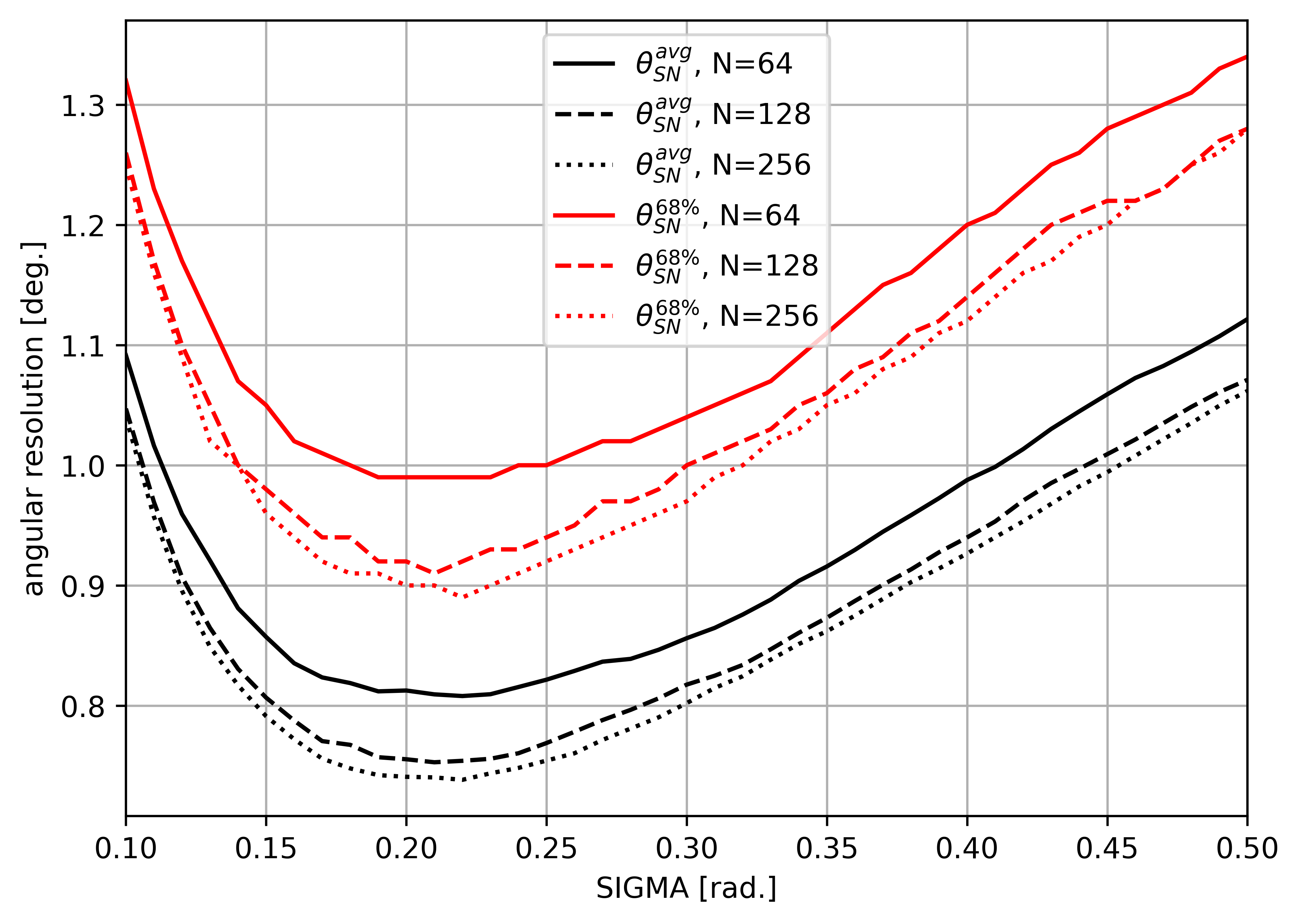}
\includegraphics[scale=0.35]{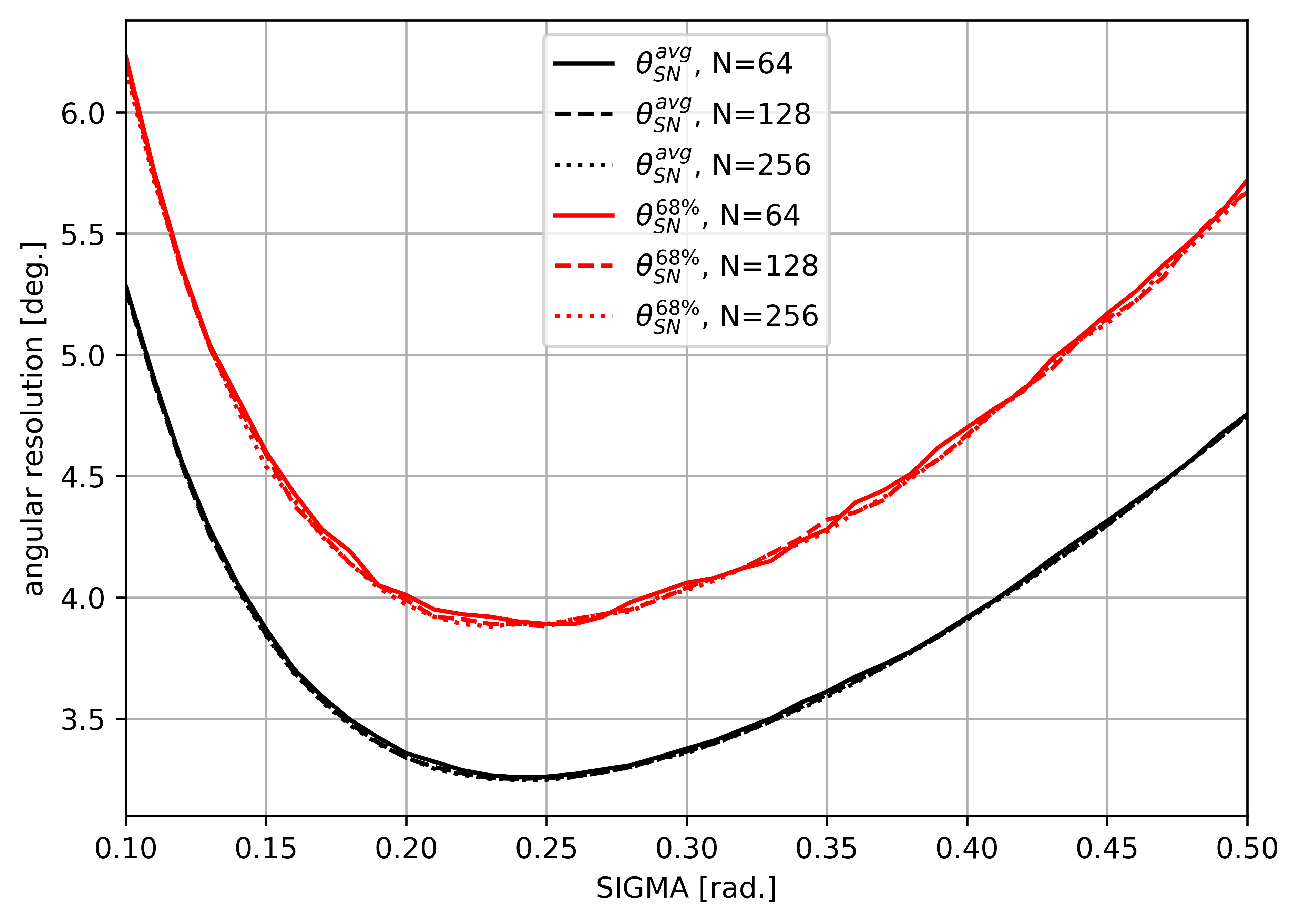}
\includegraphics[scale=0.35]{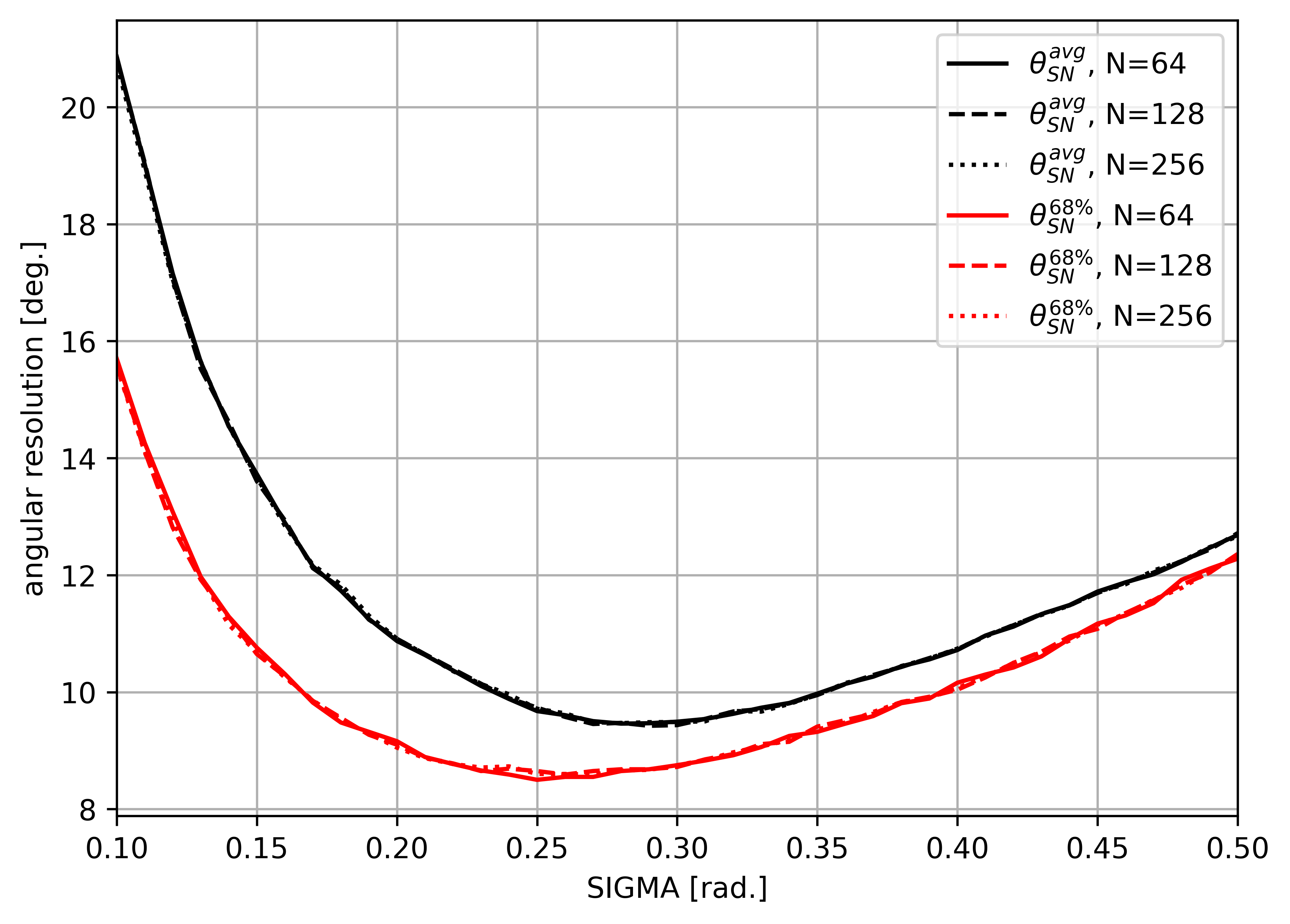}
\centering
\caption{Angular resolution vs. SIGMA for NSIDE=64, 128 and 256 for SN distance of 2 kpc (left), 10 kpc (center) and 20 kpc (right). Note the differences in scale and offset on the y-axis.}
\label{fig:ang_res_v_sigma1}
\end{figure}

To determine the values of NSIDE and SIGMA that produce the optimal angular resolution, plots of \avgdtheta\ and \dtheta{68} vs. SIGMA were created for each NSIDE for each distance. 
Figure~\ref{fig:ang_res_v_sigma1} shows the \avgdtheta\ and \dtheta{68} vs. SIGMA curves using NSIDE=64, 128 and 256 for bursts at 2 kpc (left), 10 kpc (center) and 20 kpc (right). Each curve has a minimum (corresponding to the optimal angular resolution) at a specific, optimal value of SIGMA. 
The optimal SIGMA was found for each NSIDE at every distance. 
The relative flatness of the curves in the neighborhood of the optimal SIGMA suggests that the angular resolution is not sensitive to small changes ($\pm$ 0.02 rad) in the optimal SIGMA. 

For the 2 kpc burst samples ($\sim$67\,000 events per burst), the curves for \avgdtheta\ with different NSIDE have the same shape.
For NSIDE=64, the optimal SIGMA$\approx$0.21 rad and for both the NSIDE=128 and 256, the optimal SIGMA$\approx$0.20 rad. NSIDE=128 and 256 yielded slightly better angular resolution than NSIDE=64.
Similarly, the curves for \dtheta{68} with different NSIDE had the same shape. For the NSIDE=64 curve, the optimal SIGMA$\approx$0.21 rad and for both the NSIDE=128 and 256 curves, the optimal SIGMA$\approx$0.20 rad. 
Again, NSIDE=128 and 256 gave slightly better angular resolution than NSIDE=64.

For the 10 kpc burst samples ($\sim$2700 total events per burst), the \avgdtheta\ curves for each NSIDE were nearly indistinguishable with the same optimal 
SIGMA$\approx$0.24 rad. 
For \dtheta{68}, the curves were again indistinguishable with the optimal SIGMA$\approx$0.24 rad. 

For the 20 kpc burst samples ($\sim$670 total events per burst), the curves are flatter. 
For \avgdtheta, the curves for different NSIDE are indistinguishable with the optimal SIGMA=0.29 rad.
For \dtheta{68}, the curves for different NSIDE are again indistinguishable with the optimal SIGMA=0.26 rad. 

\begin{figure}[ht!]
\includegraphics[scale=0.6]{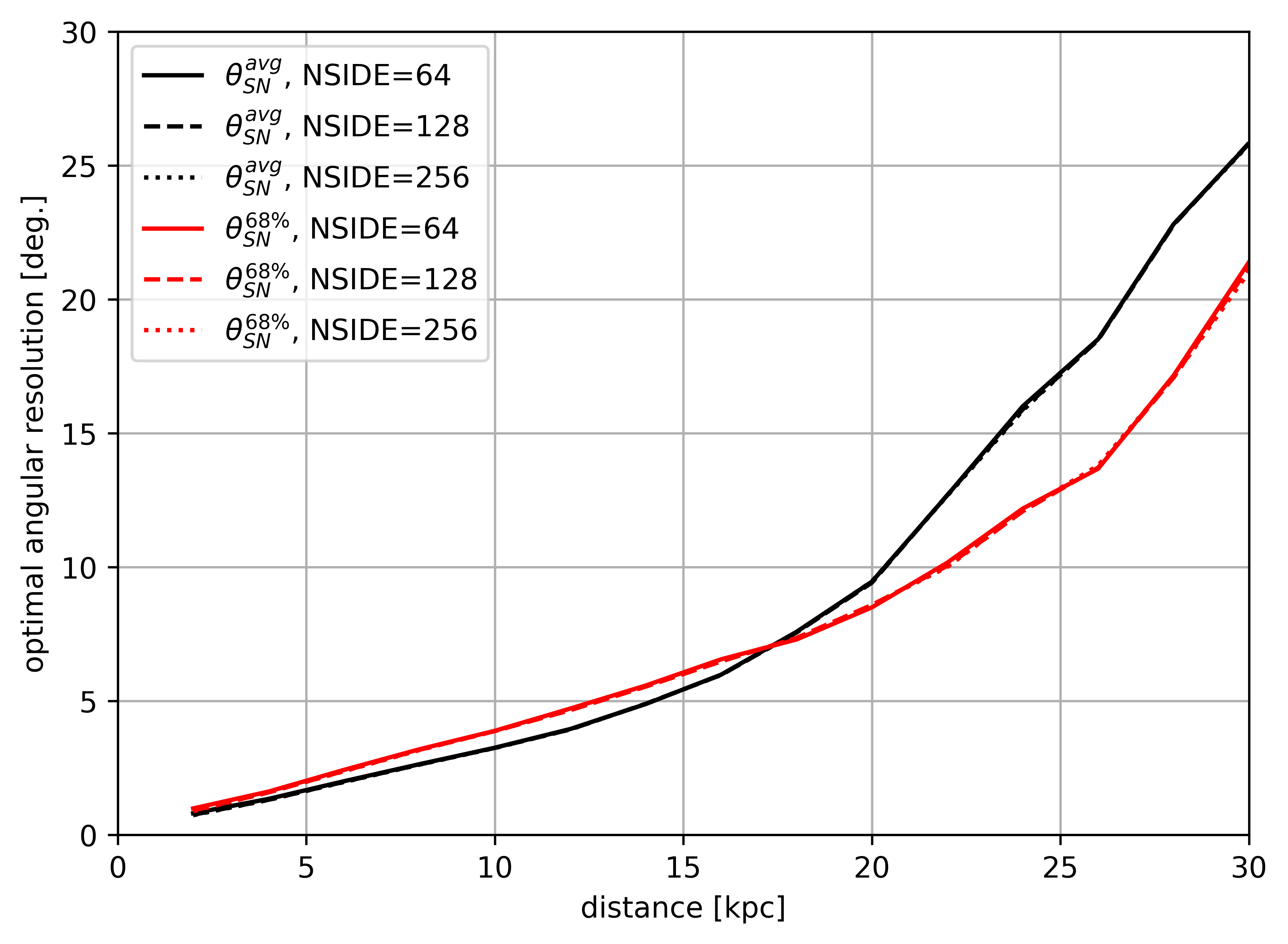}
\centering
\caption{Optimal angular resolution vs. distance using for NSIDE=64, 128 and 256 using the 
optimal SIGMA for each case. For both angular resolution measures, the curves for different values of NSIDE are indistinguishable.}
\label{fig:opt_ang_res_v_distance_all_nsides}
\end{figure}

The maximum angular resolution (\avgdtheta\ and \dtheta{68}) achievable by the HP-Fitter at each distance and for each choice of NSIDE was found using the optimal SIGMA in all cases, as shown in Figure~\ref{fig:opt_ang_res_v_distance_all_nsides}. 
For both \avgdtheta\ and \dtheta{68}, NSIDE=128 and 256 are slightly preferred at short distances but there is no significant difference at greater distances. 
Since these tests did not significantly favor one choice for NSIDE, NSIDE=128 was chosen as the default value for the HP-Fitter, mainly because it has better pixel angular resolution than NSIDE=64, but uses less total pixels than NSIDE=256.

\begin{figure}[ht!]
\includegraphics[scale=0.5]{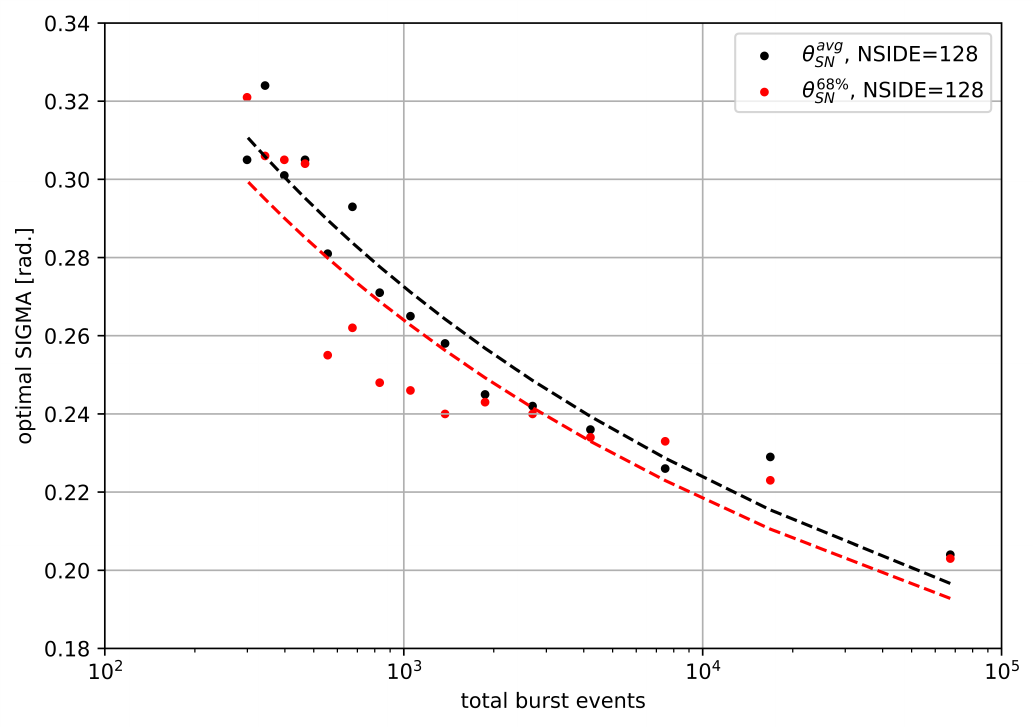}
\centering
\caption{Optimal SIGMA vs. total burst events for the \avgdtheta\ and \dtheta{68} data. The dots show the data and dashed lines show the fit the empirical equation.}
\label{fig:opt_sigma_v_ev_num}
\end{figure}

This analysis also found that the optimal SIGMA varies with distance and, therefore, depends on the total number of events per burst. 
A simple, two-parameter empirical function was used to fit the optimal SIGMA for \avgdtheta\ and \dtheta{68} vs. total burst events over the range of 300--67\,000 total burst events (corresponding to distances of 2--30 kpc).
The fit parameters from the \avgdtheta\ and \dtheta{68} data were similar, but the data for \avgdtheta\ gave a better fit. 
Therefore, the fit parameters found from the \avgdtheta\ data were chosen to use when calculating the optimal SIGMA.
Figure~\ref{fig:opt_sigma_v_ev_num} shows the optimal SIGMA for \avgdtheta\ and \dtheta{68} vs. total burst events (dots) and the dashed lines show the fits to the simple empirical model.

This optimization of the HP-Fitter was based on a simulated events generated using a single SN burst model (NK1 with NMO). Differences in SN burst neutrino characteristics might warrant changes to optimal parameters. Without additional prior knowledge of the burst event content, the directions for further optimize remain uncertain. Fortunately, HP-Fitter performance does not appear to change significantly with small changes in NSIDE or SIGMA.

\subsection{Angular Resolution vs. Distance} \label{subsec:results-ang_res}
\begin{figure}[ht!]
\includegraphics[scale=0.5]{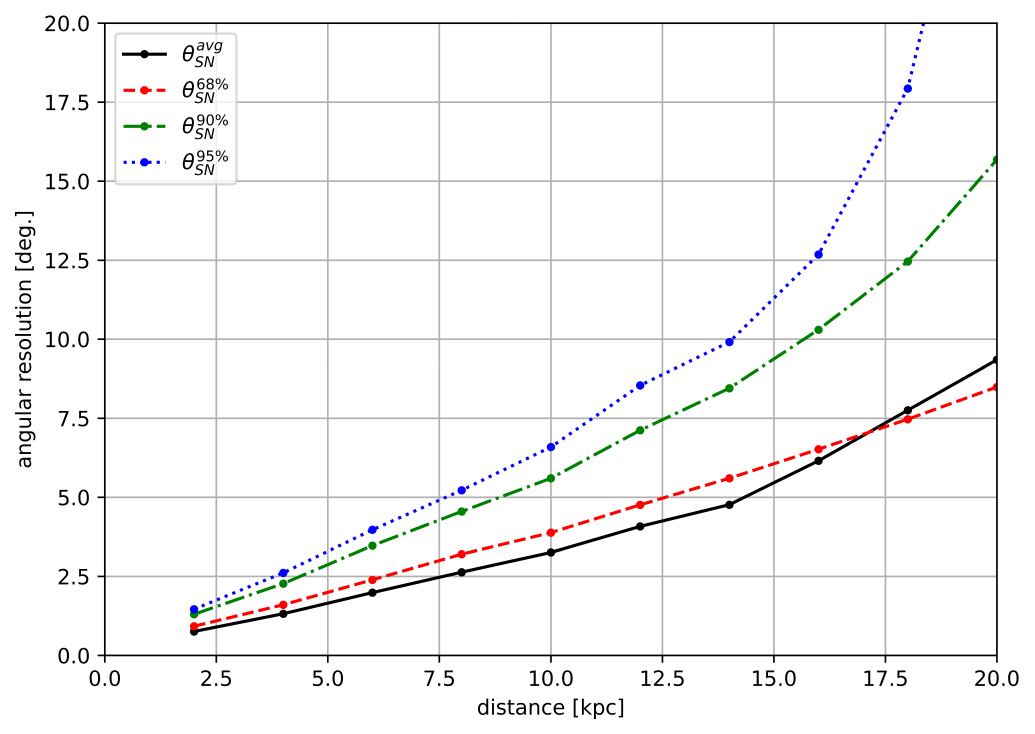}
\includegraphics[scale=0.5]{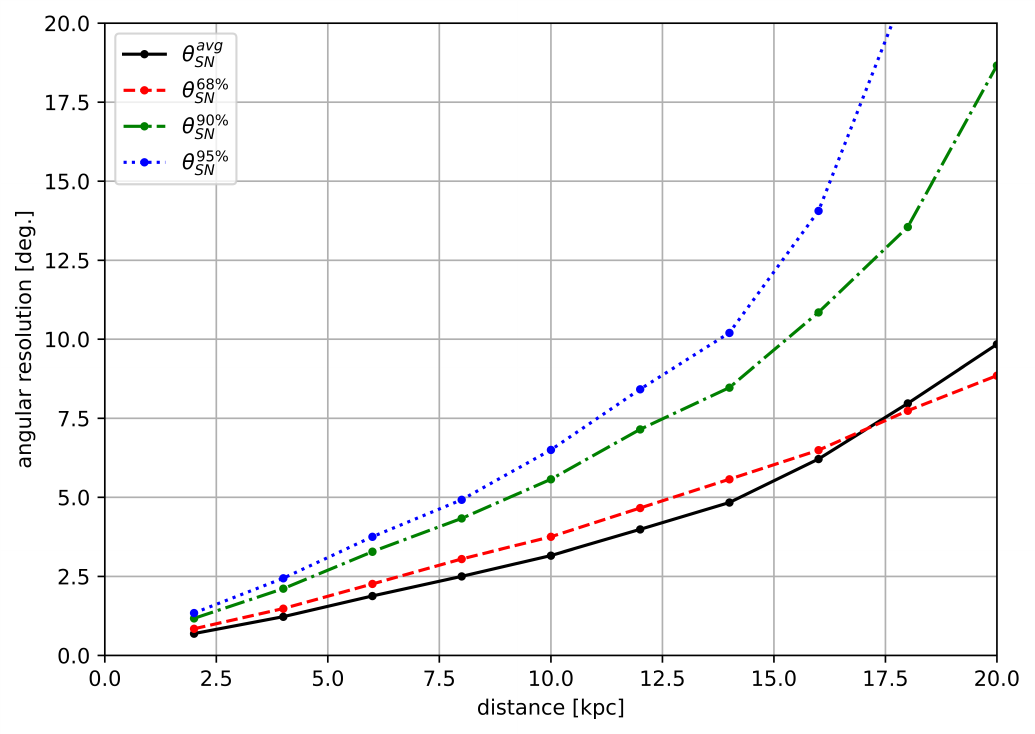}
\centering
\caption{Angular resolution measures vs. SN distance in the range of 2--20 kpc for the HP-Fitter (left) and ML-Fitter(2022) (right).} 
\label{fig:fitter_ang_res_20}
\end{figure}

The angular resolution of both fitters was measured for simulated bursts over a wide range of SN distances and, therefore, over a wide range of total burst event numbers. 
Figure~\ref{fig:fitter_ang_res_20} shows the angular resolution measures for the HP-Fitter and ML-Fitter(2022) for distances from 2--20 kpc. These values are also presented in Table \ref{tab:hp_ang_res_values} and 
Table \ref{tab:ml_ang_res_values} along with the fitter failures rates (discussed in \ref{subsec:results-fails}).

Both fitters show the same systematic changes with distance. As expected, the angular resolution worsens with increasing distance because of the decreasing number of ES events per burst. 
At small distances ($<10$ kpc), the relationship is roughly linear for all measures, suggesting that angular resolution is linearly dependent on the variance in the number of events per burst. At larger distances, \dtheta{90} and \dtheta{95} increase more rapidly, due the influence of failed reconstructions.

% results from AngResScan.NK1.202404.NMO.7.00010.00011.00013
\begin{table}[htbp]
\caption{Angular resolution measures of the HP-Fitter using NSIDE=128 for the NK1 model with NMO oscillations for SN distances from 2--20 kpc. Note that \hp pixel angular resolution for NSIDE=128 is 0.45\degree. The fitter failure rates are also listed.}
\begin{center}
\begin{tabular}{ccccccc}
\hline
distance & $\langle$events$\rangle$ & \avgdtheta & \dtheta{68} & \dtheta{90}  & \dtheta{95} & fail rate\\ 
(kpc) &  & (deg) & (deg) & (deg) & (deg) & (\%) \\ \hline
2 &    67\,350 &     0.76 &     0.92 &     1.30 &     1.46 &      0.0 \\
4 &    16\,837 &     1.32 &     1.60 &     2.27 &     2.61 &      0.0 \\
6 &     7484 &     1.99 &     2.39 &     3.47 &     3.97 &      0.0 \\
8 &     4210 &     2.63 &     3.20 &     4.55 &     5.22 &      0.0 \\
10 &     2695 &     3.25 &     3.88 &     5.60 &     6.59 &      0.0 \\
12 &     1870 &     4.08 &     4.76 &     7.12 &     8.54 &      0.1 \\
14 &     1374 &     4.76 &     5.60 &     8.45 &     9.91 &      0.2 \\
16 &     1051 &     6.15 &     6.52 &    10.30 &    12.68 &      1.1 \\
18 &      830 &     7.75 &     7.47 &    12.46 &    17.93 &      3.5 \\
20 &      674 &     9.35 &     8.49 &    15.68 &    29.62 &      5.5 \\
\hline
\end{tabular}
\label{tab:hp_ang_res_values}
\end{center}
\end{table}

\begin{table}[htbp]
\caption{Angular resolution measures of ML-Fitter(2022) for the NK1 model with NMO oscillations for SN distances from 2--20 kpc. The fitter failure rates are also listed.}
\begin{center}
\begin{tabular}{ccccccc}
\hline
distance & $\langle$events$\rangle$ & \avgdtheta & \dtheta{68} & \dtheta{90}  & \dtheta{95} & fail rate\\ 
(kpc) &  & (deg) & (deg) & (deg) & (deg) & (\%) \\ \hline
2 &    67\,350 &     0.69 &     0.84 &     1.17 &     1.34 &      0.0 \\
4 &    16\,837 &     1.22 &     1.48 &     2.11 &     2.44 &      0.0 \\
6 &     7484 &     1.88 &     2.26 &     3.28 &     3.75 &      0.0 \\
8 &     4210 &     2.50 &     3.05 &     4.33 &     4.92 &      0.0 \\
10 &     2695 &     3.16 &     3.75 &     5.57 &     6.50 &      0.0 \\
12 &     1870 &     3.98 &     4.66 &     7.15 &     8.42 &      0.1 \\
14 &     1374 &     4.83 &     5.57 &     8.47 &    10.20 &      0.2 \\
16 &     1051 &     6.21 &     6.49 &    10.85 &    14.06 &      1.1 \\
18 &      830 &     7.97 &     7.74 &    13.55 &    21.24 &      3.5 \\
20 &      674 &     9.84 &     8.85 &    18.66 &    33.78 &      5.5 \\
\hline
\end{tabular}
\label{tab:ml_ang_res_values}
\end{center}
\end{table}

\begin{figure}[ht!]
\includegraphics[scale=0.50]{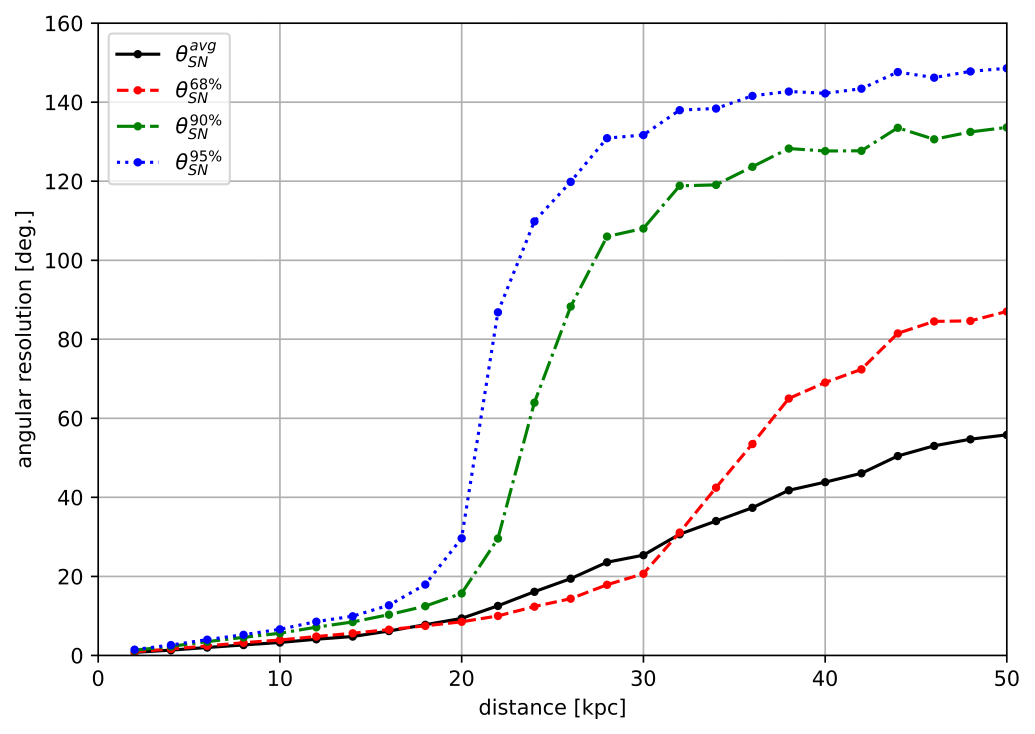}
\includegraphics[scale=0.50]{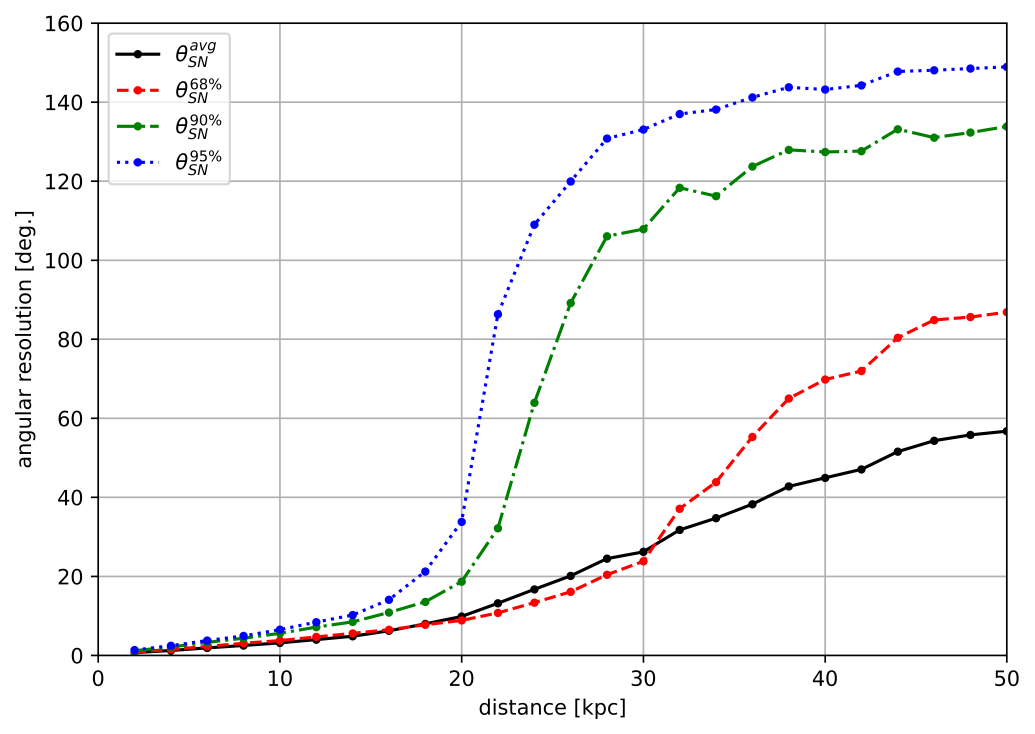} 
\centering
\caption{Angular resolution vs. distance for 2--50 kpc reconstructed by the HP-Fitter (left) and ML-Fitter(2022) (right).} 
\label{fig:fitter_ang_res_50}
\end{figure}

The impact of failed reconstructions is seen more fully in 
Figure~\ref{fig:fitter_ang_res_50}, which shows the fitter angular resolutions for distances out to 50 kpc. 
At small distances, $R_{fail}$ is negligible and the angular resolution measures are based on the ``pure'' {\thsn} distributions, as seen in Figure \ref{fig:theta_sn_anal}. 
As the SN distance increases, $R_{fail}$ also increases adding the smooth background to the {\thsn} distribution, as seen in Figure \ref{fig:vmf_fit_bgd}. 
When $R_{fail}$ passes a threshold specific to each angular resolution measure, the background in the {\thsn} distribution from the failed reconstructions causes a sharp increase in the angular resolution at a distance specific to that angular resolution measure. 

Figure~\ref{fig:fitter_ang_res_50} shows such jumps at 16, 20 and 30 kpc for \dtheta{68}, \dtheta{90} and \dtheta{95}, respectively. This demonstrates their different sensitivities to failed reconstructions. 
Beyond these distances, \dtheta{68}, \dtheta{90} and \dtheta{95} cannot accurately characterize the fitter angular resolution. 
As $R_{fail}$ approaches 100\%, \dtheta{68}, \dtheta{90} and \dtheta{95} approach their maximum values of 111\degree, 143\degree and 154\degree, respectively.
Since \dtheta{68} is valid to larger distances, it will be used exclusively as the angular resolution measure in the following analysis.

\subsection{Comparison of HP-Fitter and ML-Fitter(2022)} \label{subsec:compfitters}
\begin{figure}[ht!]
\includegraphics[scale=0.6]{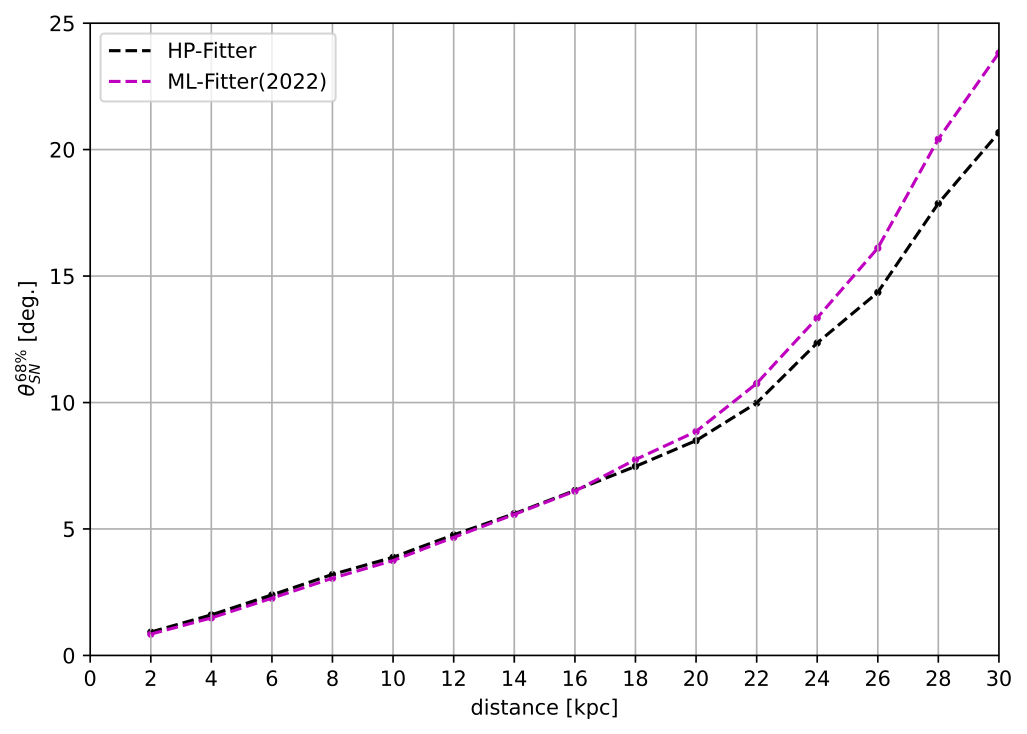}
\centering
\caption{Comparison of the \dtheta{68} angular resolutions for the HP-Fitter and ML-Fitter(2022) to distances of 30 kpc.}
\label{fig:ang_res_68_both}
\end{figure}

The angular resolutions of the HP-Fitter and ML-Fitter(2022) were very similar. 
Figure~\ref{fig:ang_res_68_both} shows the \dtheta{68} angular resolution for each fitter up to 30 kpc. For distances $\lesssim$14 kpc, ML-Fitter(2022) was marginally better, but at greater distances the HP-Fitter was superior.  

\subsection{Impact of ML-Fitter Upgrades} \label{subsec:mlfitterimproved}
\begin{figure}[ht!]
\includegraphics[scale=0.5]{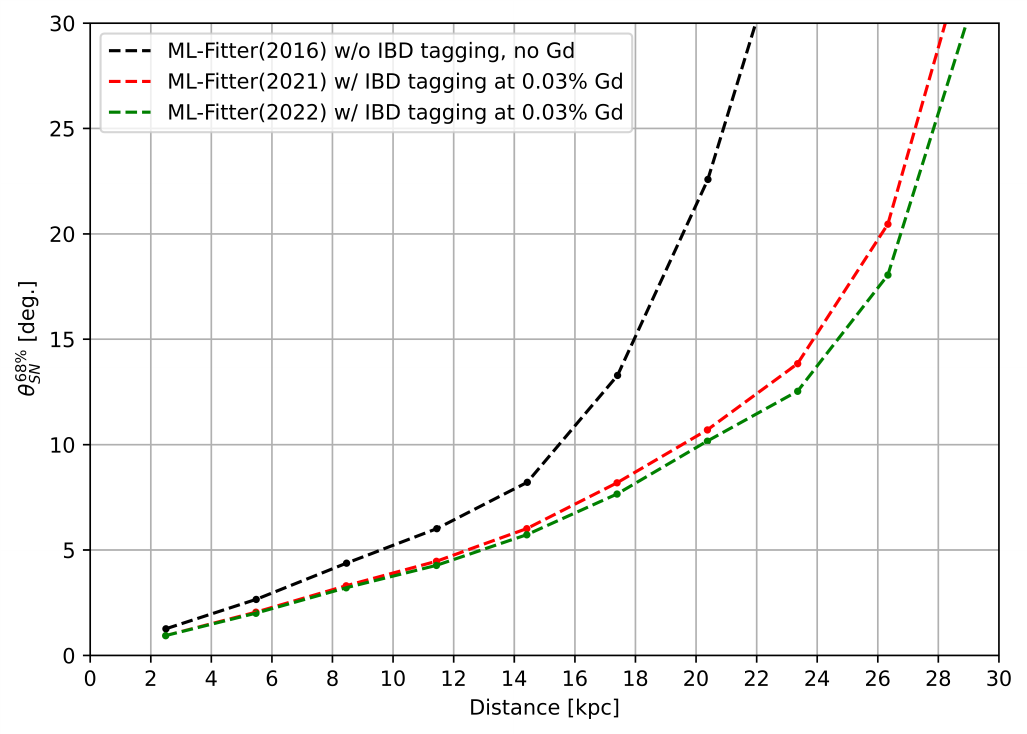}
\centering
\caption{The changes in angular resolution (\dtheta{68}) from successive upgrades of the SK ML-Fitter. Note ML-Fitter(2021) and ML-Fitter(2022) were compared based on the same Gd concentration, so differences are the results of code improvements and the use of the HP-Fitter direction.}
\label{fig:mlfitter_progress}
\end{figure}

The original SK ML-Fitter(2016) was upgraded to ML-Fitter(2021) by using IBD-tagging information to reduce the non-ES background. ML-Fitter(2022) incorporated other further changes, including using the HP-Fitter for the initial values for the SN direction fit parameters. 
Figure~\ref{fig:mlfitter_progress} shows impact of the successive upgrades on angular resolution. 
The inclusion of IBD tagging information in ML-Fitter(2021) yielded the largest improvements. 
The upgrades in ML-Fitter(2022) yielded only a small improvement in angular resolution, but had a significant impact on reconstruction times (see \ref{subsec:fitterspeedmethod}).

\subsection{Impact of Gd Concentration} \label{subsec:mlfitter_gd_conc}
\begin{figure}[ht!]
\includegraphics[scale=0.5]{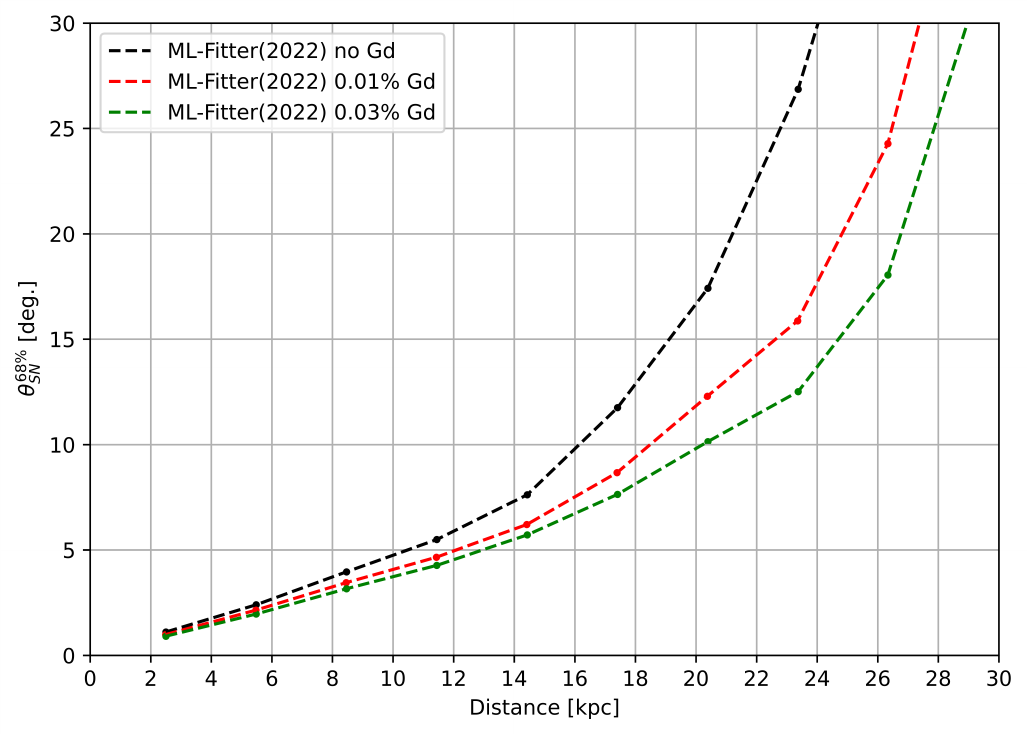}
\centering
\caption{A comparison of the fitter angular resolution for different Gd-loading.}
\label{fig:mlfitter_gd}
\end{figure}

The gains in ML-Fitter performance from using IBD tagging information should depend on the Gd concentration, which determines the neutron-capture detection efficiency. 
Figure~\ref{fig:mlfitter_gd} shows the angular resolution for the same fitter, ML-Fitter(2022), with different Gd loadings 
(0\% for before SK-Gd, 0.01\% for SK-VI and 0.03\% for SK-VII). 
The successive increases in Gd concentration show significant improvements in angular resolution, especially at greater SN distances.

\subsection{Failure Rates} \label{subsec:results-fails}
\begin{figure}[ht!]
\includegraphics[scale=0.5]{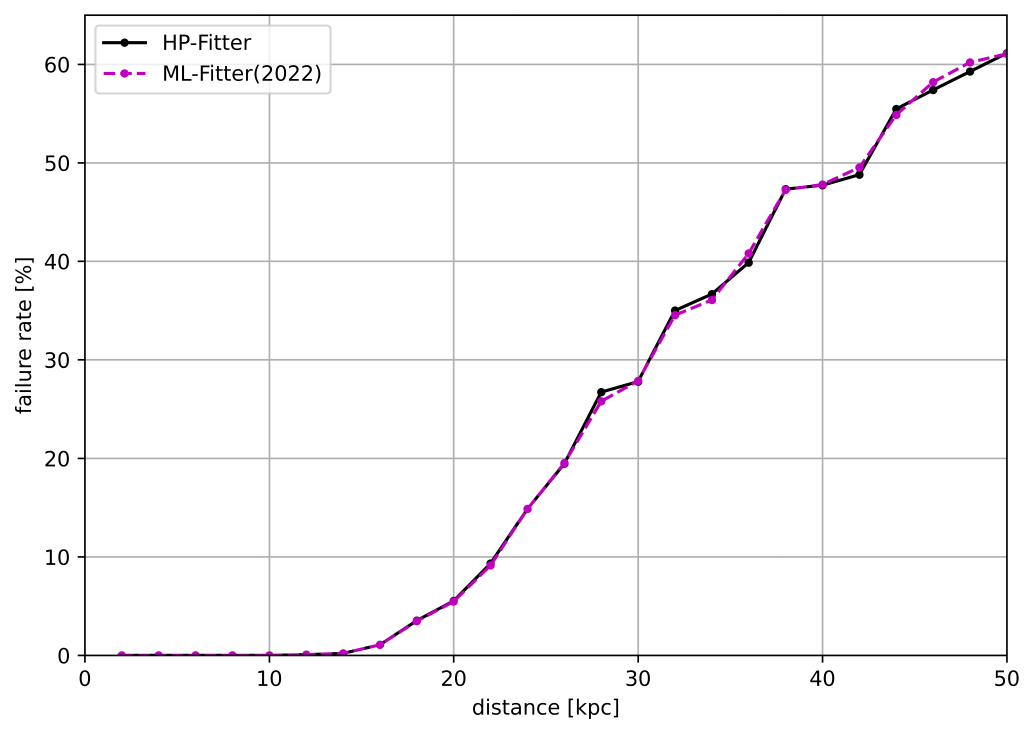}
\centering
\caption{The reconstruction failure rates vs. distance to 50 kpc for the HP-Fitter and ML-Fitter(2022).}
\label{fig:both_fitter_failrate_curves}
\end{figure}

Figure~\ref{fig:both_fitter_failrate_curves} shows the estimated failure rates of the HP-Fitter and ML-Fitter(2022) to distances of 50 kpc. 
The non-smooth features are artifacts of the estimation method.
The curves are nearly identical. For distances $\lesssim$10 kpc, the failure rate is negligible for both fitters. The failure rates increase to $\sim$5\% at 18 kpc and reach $\sim$40\% at 35 kpc. 

\subsection{Angular Resolution and Failure Rate Matrices} \label{subsec:resmatrix}
The above analysis characterizes the fitter performance over a wide range of total burst events. However, it is based on a single SN burst model, so the relative number of ES and non-ES events per burst, which also impacts direction reconstruction, is fixed.

In order to assess fitter performance over a wide range of both relative and absolute ES and non-ES events per burst, 
a set of burst samples was generated with independent numbers of ES and non-ES events.
Non-ES events per burst ranged from 0 to 70\,000 event in 16 discrete bins. 
ES events per burst ranged from 0 to 3\,000 events in 16 discrete bins. 
The different combinations span a 16 $\times$ 16 matrix with total of 256 different types of burst sample.
For example, the burst samples in the first bin have 0--1000 non-ES events and 0--25 ES events per burst.
For each burst type (corresponding to a bin in the matrix), 3000 burst samples were generated by randomly drawing the appropriate range of ES and non-ES events from the event pool. Each burst sample was then processed by each fitter to calculate \dnusnrecon. The ML-Fitter also calculates the number of ES and non-ES events in the burst.

When all the samples for the matrix were processed, they were sorted into bins according to the number of ES and non-ES events per burst. For the HP-Fitter, binning was based on the true event numbers. For the ML-Fitter, the bursts could be sorted based on the true event numbers or the calculated (``fit'') event numbers.
For each fitter, the calculated \dnusnrecon for all bursts in a bin were used to construct a characteristic \thsn\ distribution from which the angular resolution and failure rate were extracted.
The angular resolution values for each forms an ``angular resolution matrix'' and the failure rate values for each bin forms a ``failure rate matrix''. For the HP-Fitter, only matrices binned using true event numbers can be made. For the ML-Fitter, matrices binned using true or fit event numbers can be made. 
Angular resolution and failure rate matrices binned by fit event numbers are useful because they are based on the same information calculated by the ML-Fitter when a SN neutrino burst is detected. In the event of a real burst, these matrices could be referenced to provide estimates of the angular error and failure likelihood of the reconstructed pointing information.

In testing, it was found that using a simple method to sample ES and non-ES events is undermined by the correlations between different event types, making it unphysical to treat ES and non-ES event numbers as independent variables. 
Since the numbers and energy distributions of IBD ($\bar{\nu}_e + p \rightarrow e^+ + n$) and ES2 ($\bar{\nu}_e + e^- \rightarrow \bar{\nu}_e + e^-$) events both depend on the luminosity and energy distribution of $\bar{\nu}_e$ in the SN neutrino burst, the IBD and ES2 event numbers in a burst cannot be varied independently. 
In fact, this relationship is used in the ML-Fitter to reduced the number of fit parameters 
(see Appendix \ref{sec:mlfitterdetails}).
This is also true for the O16CC2 ($\bar{\nu}_e + {\rm ^{16}O} \rightarrow e^+ + X$) events, but these are a much smaller component of the non-ES events so the impact is less significant.
Likewise, O16CC1 ($\nu_e + {\rm ^{16}O} \rightarrow e^- + X$) and ES1 ($\nu_e + e^- \rightarrow \nu_e + e^-$) events both depend on the luminosity and energy distribution of $\nu_e$ in the SN neutrino burst. 
Since O16CC1 is also a small component of the non-ES events, these the impact of these correlations is less significant.

To reduce the effects of these correlations, the ES events from different reaction channels are treated differently. 
The ES1 + ES3 + ES4 (=``ES$^*$'') events are treated as being independent of the non-ES events.
To generate a burst sample, non-ES events are drawn randomly from the event pool using the non-ES range for that bin. A sample of ES$^*$ events is also drawn randomly from the event pool based on the ES number range for that bin. 
To these are added a sample of ES2 events with a distribution based on the IBD event distribution in the non-ES sample. 

\begin{figure}[htbp]
\includegraphics[scale=0.55]{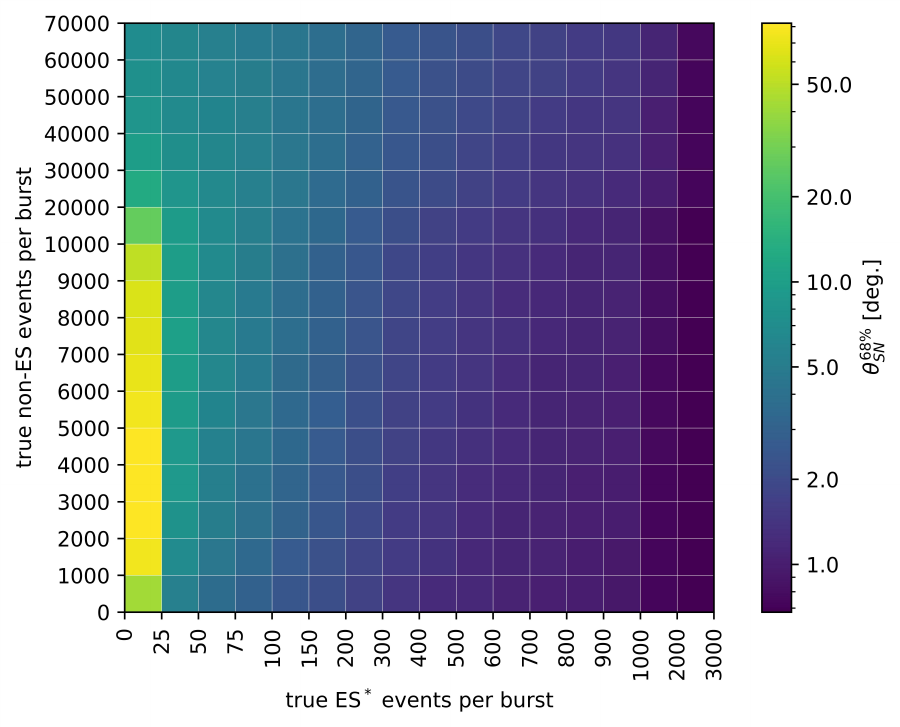}
\includegraphics[scale=0.55]{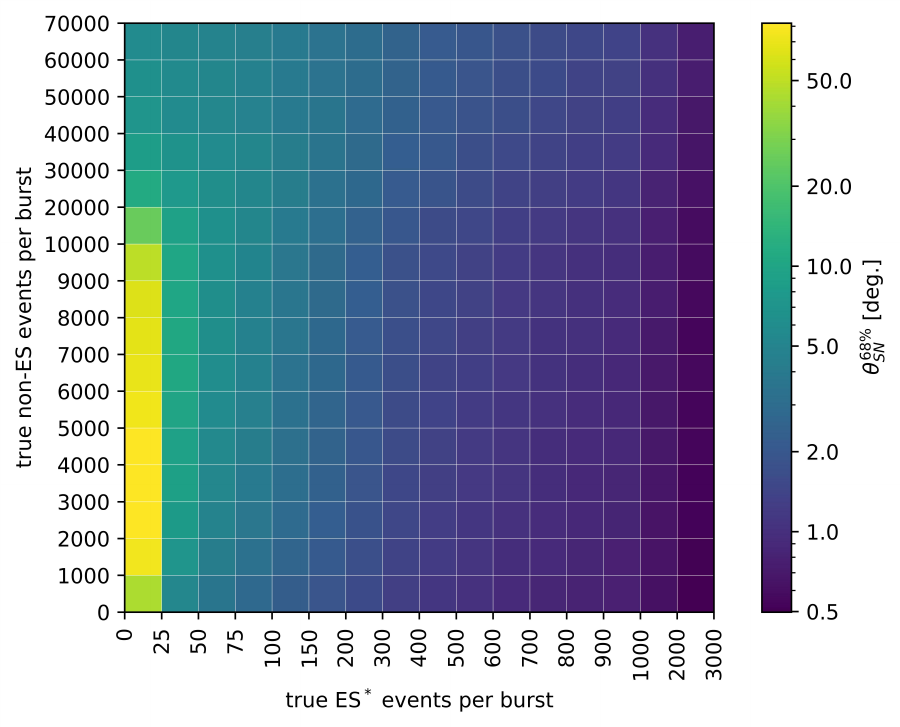}
\centering
\caption{The \dtheta{68} angular resolution matrices for the HP-Fitter (left) and ML-Fitter(2022) (right) binned based on true event numbers. A log scale is used to enhance the differences at low values.}
\label{fig:comp_ang_res_matrix_hp_ml_true_bins}
\end{figure}

Figure~\ref{fig:comp_ang_res_matrix_hp_ml_true_bins} shows the \dtheta{68} angular resolution matrices for the HP-Fitter and ML-Fitter(2022). 
Both are binned based on the true number of ES$^*$ and non-ES events (IBD + O16CC) per burst. 
The results for both fitters are comparable over all ranges of non-ES and ES$^*$ event numbers. 
As expected, both have the worst angular resolution for bursts with low absolute and/or relative numbers of ES$^*$ events (first two columns). 
For most other bins, the angular resolution is \dtheta{68} $<10$\degree. 

\begin{figure}[ht!]
\includegraphics[scale=0.55]{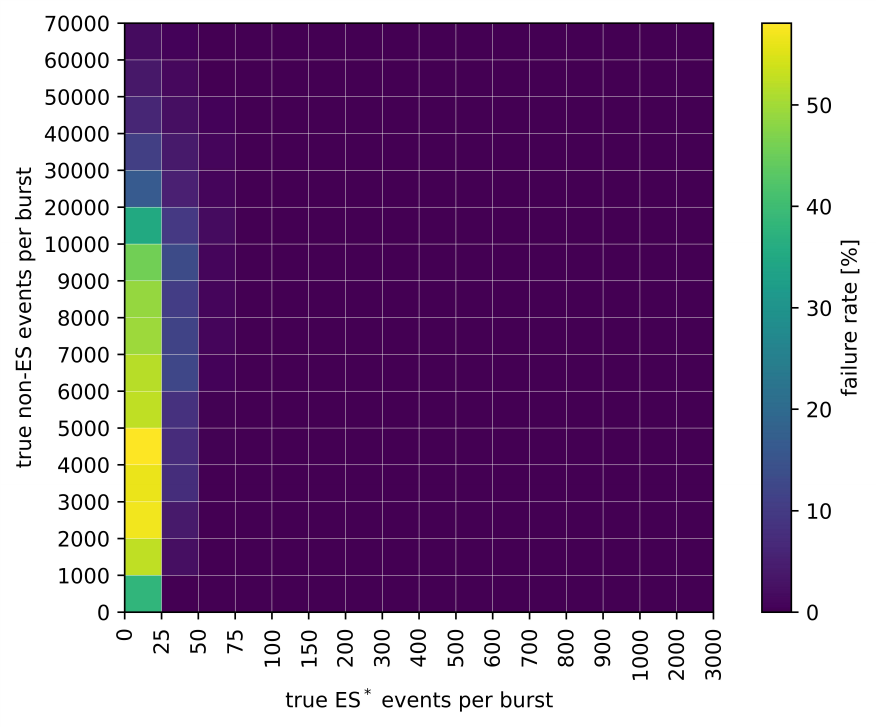}
\includegraphics[scale=0.55]{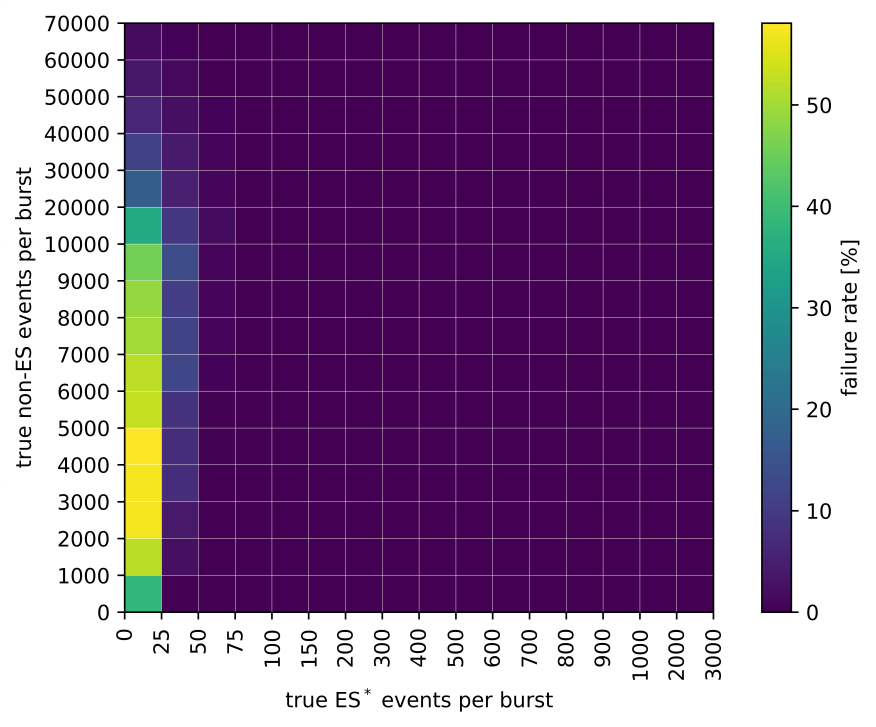}
\caption{Fitter reconstruction failure rates for the HP-Fitter (left) and the ML-Fitter(2022) (right) binned according to the true event numbers.} 
\label{fig:fail_matricesl}
\centering
\end{figure}

Similarly, failure rate matrices characterize the fitter reconstruction failure over a wide range of absolute and relative ES events per burst. 
Figure~\ref{fig:fail_matricesl} shows the failure rate matrices for the HP-Fitter and ML-Fitter(2022), both with the burst samples binned based on true event numbers. 
The failure rate matrices for both fitters show nearly identical results. The failure rates are very low, except when the absolute numbers of ES events are very small (first column).

% \hspace{8pt}

\begin{figure}[htbp]
\includegraphics[scale=0.50]{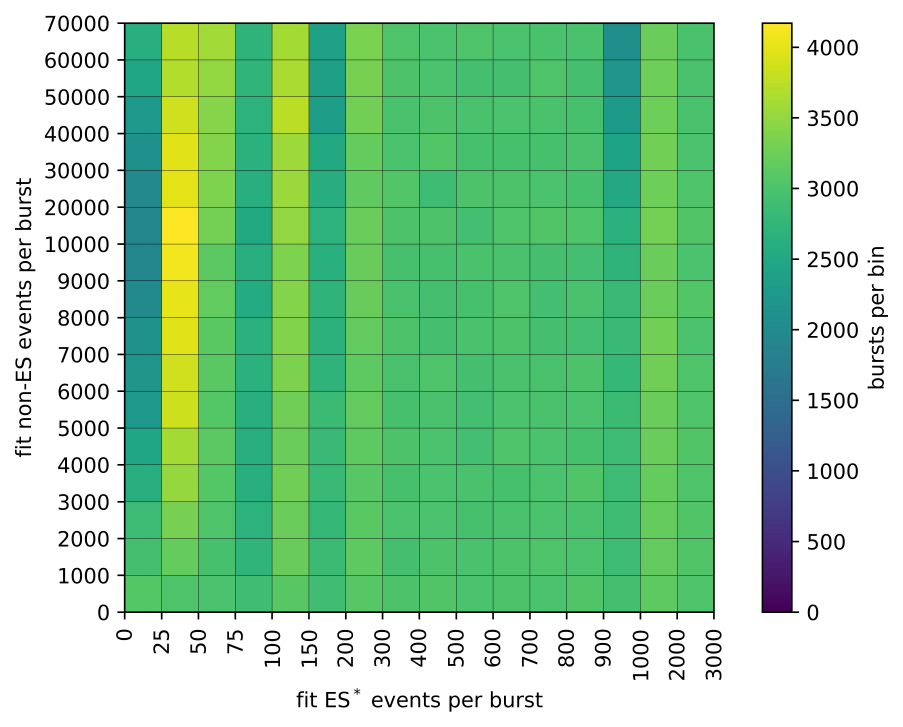}
\centering
\caption{Matrix showing the number of burst samples per bin based on ML-Fitter fit event numbers where the true number of bursts per bin was 3000.}
\label{fig:comp_matrix_binning}
\end{figure}

Binning the bursts based on the ``fit'' non-ES and ES event numbers calculated by the ML-Fitter causes some distortions in matrices because some bursts are added to the wrong bin. 
Figure~\ref{fig:comp_matrix_binning} shows the number of bursts per bin when burst samples binned based on the fit event numbers from the ML-Fitter. 
Each bin should contain 3000 bursts, but some are displaced due to errors in the ML-Fitter calculated event numbers. This is most pronounced in bursts with low numbers of ES events. 

\begin{figure}[htbp]
\includegraphics[scale=0.55]{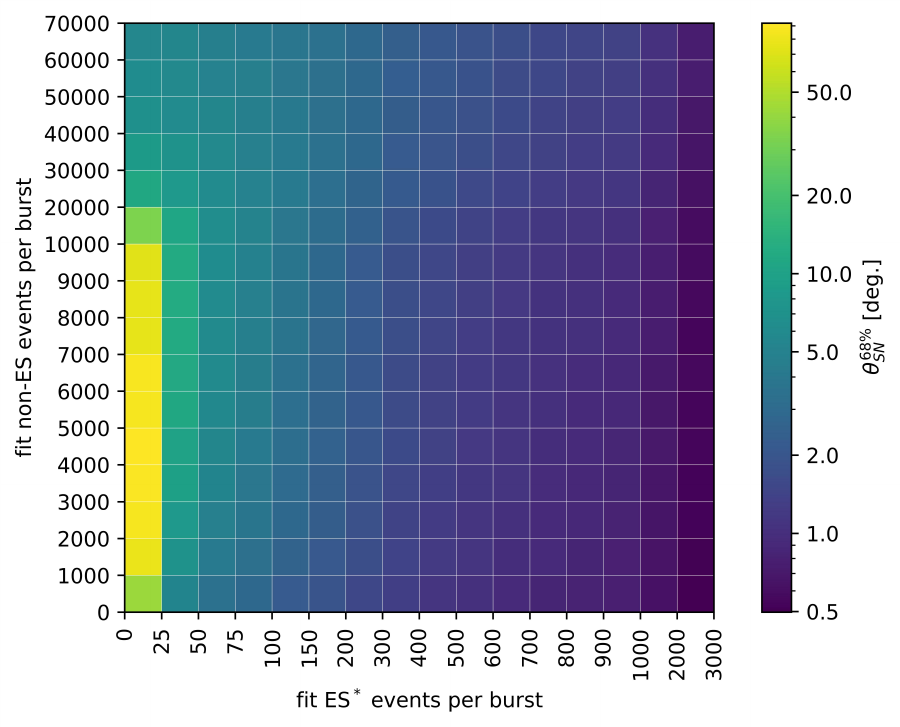}
\includegraphics[scale=0.55]{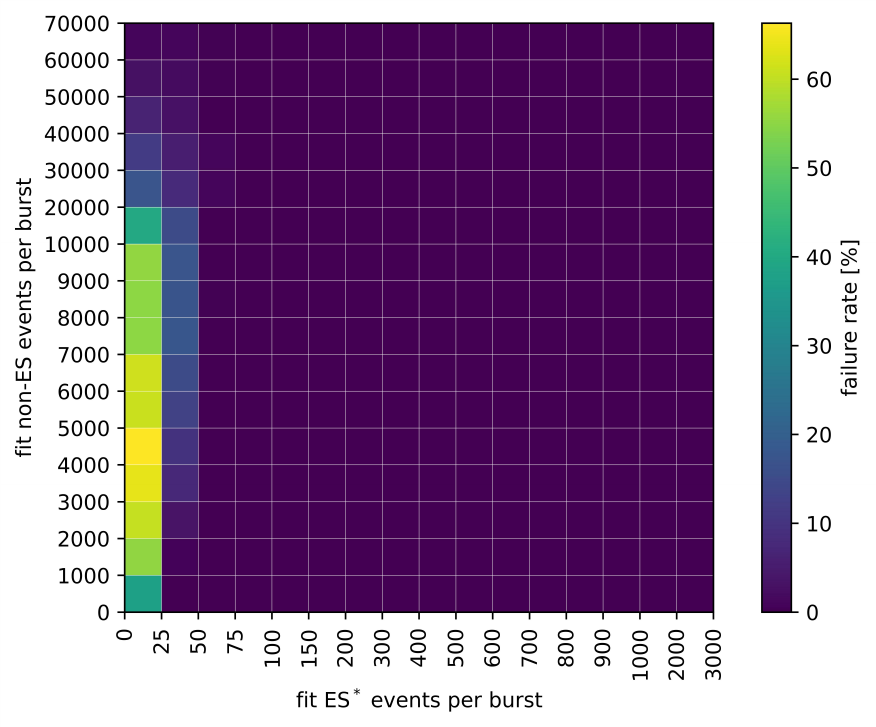}
\centering
\caption{Angular resolution (\dtheta{68}) (left) and failure rate (right) matrices for ML-Fitter(2022) binned using non-ES and ES event numbers calculated by the fitter. These values may be used to estimate the localization accuracy in the event of a SN alert.}
\label{fig:comp_ang_res_matrix_ml_true_fit_bins}
\end{figure}

Figure~\ref{fig:comp_ang_res_matrix_ml_true_fit_bins} shows the angular resolution and failure rate matrices for the ML-Fitter binned using fit event numbers. 
These results show only minor differences from those for the ML-Fitter binned with true events shown in Figure~\ref{fig:comp_ang_res_matrix_hp_ml_true_bins} and Figure~\ref{fig:fail_matricesl}.
In the event of SN alert, the values in these matrices could be used in SNWATCH to provide estimates the angular error and probability of failure in SN localization based on the fit number of non-ES and ES2 events numbers. 

\subsection{Direction Reconstruction Speed} \label{subsec:fitterspeedmethod}
The speed of the new fitters was tested by measuring the computational time over thousands of burst samples. 
This did not include the computational time for event reconstruction, SN burst identification, alert generation or other ancillary functions. 
For the HP-Fitter and ML-Fitter(2022) the reconstruction times were measured for calculations done with a Windows desktop PC, without GPU acceleration\footnote{13th Gen Intel Core i7-13700F, 2.10 to 5.20 GHz, 16 cores, 32 GB RAM.} 
For ML-Fitter(2021), the reconstruction times were estimated from calculations performed on \snww computers.
The direction reconstruction times for different fitters are compared in Table~\ref{tab:fitterspeeds}. 

\begin{table}[htbp]
\centering
% \begin{table}[htbp]
\caption{The approximate direction reconstruction times for various fitters.}
% \begin{center}
\begin{tabular}{cccc}
\hline
Fitter & \multicolumn{3}{c}{Direction Reconstruction Times (sec.)} \\
& $\sim$3000 events & $\sim$10\,000 events & $\sim$60\,000 events \\
\hline
ML-Fitter(2016) & - &  - &513 \\
ML-Fitter(2021) & 19 &  63 & 376 \\
HP-Fitter  & 0.4 & 0.4 & 0.4 \\
ML-Fitter(2022) & 1.0 & 2.2 & 11.1 \\
\hline
\end{tabular}
\label{tab:fitterspeeds}
\end{table}

The new fitters show a remarkable improvement in speed. The HP-Fitter direction reconstruction times are negligible and essentially independent of the number of burst events. ML-Fitter(2022) is significantly faster than the earlier C++ based fitters, but the reconstruction times increase with the number of events per burst. The reconstruction times for both new fitters are negligible compared to the time required for event reconstruction, $\mathcal{O}$(min).

\section{Conclusion and Prospects} \label{sec:summary}
The inclusion of IBD-tagging information from SK-Gd and the development of new and improved direction fitters have significantly advanced the capabilities of SNWATCH at SK.  
The availability of faster and more accurate SN localization for SN early-warning alerts significantly increases the opportunity for early multi-messenger observations of the next nearby SN. 

The new HP-Fitter is a novel method that uses a \hp sphere as a data structure to map and analyze the 3-d angular distribution of the reconstructed event directions. Neutrino-electron elastic scatter events create an ES-peak in the smoothed \hp burst event map that is co-directional to the SN neutrino wavefront.
The HP-Fitter is inherently fast with angular resolution and failure rates comparable to those from the more complex ML-Fitter. 
This method might be useful for other applications in particle physics. 

In addition, the \snww maximum-likelihood fitter methodology underwent several upgrades. The use of IBD-tagging information was found to significantly improve the angular resolution. Changes to the code yielded major reductions in computation time. 

The new and improved direction fitters have been tested and are now integrated into SNWATCH. Currently they are used in tandem, with the HP-Fitter providing the initial SN direction estimate for ML-Fitter(2022), which calculates the SN pointing direction. It also calculates the number of events per reaction channel which may be used by the angular resolution and failure rate matrices to provide estimates of the angular error and likelihood of fitter failure.

The improvements to \snw, along with changes in the SN response policies and procedures at SK, 
have important implications for the optimal strategies for multi-messenger observer responses to an SN alert. 
Previously, SK was known to have technical and operational bottlenecks that could result in a long latency, $\mathcal{O}$(hrs),  
between the detection of an SN burst detection at SK and the public release of detailed SN information, including localization. 
SK has recent implemented new, GCN-based, SN alert system, ``SK\_SN''\footnote{https://gcn.nasa.gov/missions/sksn}. 
In the event of an SN, \snww will issue a machine-readable GCN notice which will include information on detector status, number of burst events, discovery time and localization information from the new 
fitters\footnote{The NASA GCN website, https://gcn.nasa.gov, provides detailed documentation and instructions for implementing ``listener'' monitoring of GCN notices including SK\_SN.}.
Currently, a SK\_SN GCN notice can be issued within $\sim$90 seconds following burst detection, and work continues to further reduce this latency.

The availability of this new, lower-latency SN alert with accurate pointing information gives the multi-messenger astronomical community an unprecedented opportunity to observe the earliest emissions from a nearby core-collapse SN. 
Potential observers might benefit from reviewing their SN alert response plans.
Several issues need to be considered, including:
1) the priority of a galactic SN alert in the target of opportunity (ToO) policies and procedures,
2) operational changes to minimize the latency between receiving the alert, ending current observing sessions and the changing the instrument configuration (filters, stops, slewing, etc.) 
to prepare to search for the SN, 
3) coordination between observers with different local conditions to provide optimal coverage of the target region, including tiling parameters, and ensuring coverage over the uncertain appearance time.
In the event, the likelihood of observing the initial SBO depends on minimizing the response time, especially if the shock propagation time is short. 
Optimal response plans vary between instruments based on their capabilities. For an example, \cite{andreoni2024rubin2024envisioningvera} describes a proposed ToO response plan for a galactic SN alert for the Vera C. Rubin Observatory, including the role of the SK pointing direction. 

Development continues on SNWATCH to find new ways to improve the pointing accuracy and reduce the alert latency. These includes improving the speed and accuracy of event reconstruction, burst identification and IBD event tagging. In addition, the use of lower energy thresholds and further optimization of HP-Fitter and ML-Fitter parameters, methodologies and code are being investigated. 

The continuing goal of this work is to support the global, multi-messenger astronomical community as we all prepare for, and anticipate, the spectacle of the next galactic supernova.

\begin{acknowledgments}
\input{SK-paper-acknowledgements-20251121.tex}

\end{acknowledgments}

\software{
    Astropy~\citep{2013A&A...558A..33A,2018AJ....156..123A},  
    Healpy~\citep{2005ApJ...622..759G}, 
    iMinuit~\citep{iminuit},
    NumPy~\citep{harris2020array},
    matplotlib~\citep{Hunter:2007},
    SciPy~\citep{2020SciPy-NMeth},
    Uproot~\citep{jimpivarski20203952728},
    SNEWPY~\citep{Baxter2021, Baxter_2022}
    }

\appendix
\section{ML-based Direction Reconstruction at SK} \label{sec:mlfitterdetails}
The original SN direction reconstruction method for SNWATCH, ML-Fitter(2016), ~\citep{Abe:2016waf} uses maximum likelihood methods to find values of SN direction and event numbers per reaction channel in coarse energy bins. 

In this implementation the events are binned into four reaction channels (``$r$-bins'') and five coarse energy bins (``$k$-bins'') with characteristics listed in Table \ref{tab:mlfitparams}. These bins form a $4 \times 5$ array, $N_{r,k}$, where $\sum_{r,k} N_{r,k}$ = the total number of events. 

\begin{table}[htbp]
\caption{Neutrino reaction channels used in SK SN Direction ML-Fitter.}
\begin{center}
\begin{tabular}{clccc}
\hline
$r$-bin & Reaction Channel & & $k$-bin & Energy Range (MeV)\\ \hline
0 & IBD: & & 0 & $7 < E < 10$ \\ 
1 & ES2: & & 1 & $10 < E < 15$ \\
2 & ES$^*$ (= ES1+ES3+ES4): & & 2 & $15 < E < 22$ \\
3 & O16 (= O16CC1+O16CC2): & & 3 & $22 < E < 35$ \\
 & & & 4 & $35 < E < 50$ \\
\hline
\end{tabular}
\label{tab:mlfitparams}
\end{center}
\end{table}

This event binning scheme would require fitting 20 event number parameters in addition to the two SN angles for a total of 22 fit parameters.
A few symmetries and approximations were exploited to reduce the number of fit parameters. First, the ES2 events per $k$-bin, $\vec{N}_{1,k}$, can be calculated from the IBD events per $k$-bin, $\vec{N}_{0,k}$, using a $5 \times 5$ transformation matrix, $\hat{A}_{1,k,~0,k}$, because both depend on the $\bar{\nu}_e$ fluence and spectrum. This removes the need to independently fit the five ES2 event number parameters.

\begin{equation}
\vec{N}_{1,k} = \hat{A}_{1,k,~0,k} \cdot \vec{N}_{0,k} \:
\label{eq:transform}
\end{equation}

In addition, the original implementation of the ML-Fitter assumes that the number of O16CC events with $E<22$ MeV is negligible, so the $N_{3,k=0,1,2}$ are fixed to zero and only $N_{3,k=3,4}$ are free fit parameters.

These simplifications reduce the number of fit parameters in $N_{r,k}$ to 12 and with the two direction angles, for a total of 14 fit parameters to be found by ML optimization.

The extended likelihood for individual events, $L_i$, is calculated by:
\begin{equation}
L_{i} = \sum_{r} N_{r,k_i} p_{r}(E_i, \theta_i) \:
\label{eq:likelihood2}
\end{equation}

where $p_{r}(E_i, \theta_i)$ is the probability density function (PDF) for events in reaction channel, $r$, having a energy, $E_i$, and scattering angle, $\theta_{i}$ (=$\hat{d}_i \cdot \hat{d}_{SN}$) where $\hat{d}_i$ and \dnusn are the direction vectors of the outgoing particles and SN neutrino wavefront, respectively. Since the direction of the SN neutrino wavefront is opposite to the sky location of the SN (\dsntrue\ = $-$\dnusn), the reconstructed pointing direction to the SN is found as \dsnrecon = $-$\dnusnrecon. 

In the current version of the ML-Fitter, the PDFs for IBD ($r$=0) and O16 ($r$=3) events are calculated from simple empirical equations that are functions of $E_i$ and $\cos \theta_{i}$.
For ES$^*$ reaction channel events ($r$=2), PDFs are calculated used empirical equations based on fits of MC simulated PDFs of these events that depend on $E_i$ and $\cos\theta_{i}$.

The total extended-likelihood for all events in the burst is then:
\begin{equation}
\mathcal{L} = \exp(-\sum_{r,k} N_{rk}) \prod_{i} L_i
\label{eq:fulllikelihood2}
\end{equation}

The initial values for the SN direction angle parameters is found using a coarse grid search. The event directions are binned according to a 2-d grid of fixed azimuthal and polar angles. The angles of the bin with the most events are used as initial search values for the SN direction parameters for the fit. 

The initial values for the event number parameters are found by first estimating the number of non-ES events from the number of large angle scattered events per $k$-bin. This is then used to estimate the number ES-event per $k$-bin.

The total likelihood function is then optimized using Minuit to find the SN direction angles and the number of events per reaction channel by:
\begin{equation}
\pdv{\mathcal{L}}{N_{r,k}} = \pdv{\mathcal{L}}{\hat{d}_{SN}} = 0
\label{eq:optlikelihood2}
\end{equation}

\bibliography{sksndirection}{}
\bibliographystyle{aasjournal}

%TC:endignore

\end{document}

%% file: authors-merge-barry-orcid2.tex
\newcommand{\AFFicrr}{\affiliation{Kamioka Observatory, Institute for Cosmic Ray Research, University of Tokyo, Kamioka, Gifu 506-1205, Japan}}
\newcommand{\AFFkashiwa}{\affiliation{Research Center for Cosmic Neutrinos, Institute for Cosmic Ray Research, University of Tokyo, Kashiwa, Chiba 277-8582, Japan}}
\newcommand{\AFFipmu}{\affiliation{Kavli Institute for the Physics and
Mathematics of the Universe (WPI), The University of Tokyo Institutes for Advanced Study,
University of Tokyo, Kashiwa, Chiba 277-8583, Japan }}
\newcommand{\AFFmad}{\affiliation{Department of Theoretical Physics, University Autonoma Madrid, 28049 Madrid, Spain}}
\newcommand{\AFFubc}{\affiliation{Department of Physics and Astronomy, University of British Columbia, Vancouver, BC, V6T1Z4, Canada}}
\newcommand{\AFFbu}{\affiliation{Department of Physics, Boston University, Boston, MA 02215, USA}}
\newcommand{\AFFuci}{\affiliation{Department of Physics and Astronomy, University of California, Irvine, Irvine, CA 92697-4575, USA }}
\newcommand{\AFFcsu}{\affiliation{Department of Physics, California State University, Dominguez Hills, Carson, CA 90747, USA}}
\newcommand{\AFFcnm}{\affiliation{Institute for Universe and Elementary Particles, Chonnam National University, Gwangju 61186, Korea}}
\newcommand{\AFFduke}{\affiliation{Department of Physics, Duke University, Durham NC 27708, USA}}
\newcommand{\AFFgifu}{\affiliation{Department of Physics, Gifu University, Gifu, Gifu 501-1193, Japan}}
\newcommand{\AFFgist}{\affiliation{GIST College, Gwangju Institute of Science and Technology, Gwangju 500-712, Korea}}
\newcommand{\AFFuh}{\affiliation{Department of Physics and Astronomy, University of Hawaii, Honolulu, HI 96822, USA}}
\newcommand{\AFFicl}{\affiliation{Department of Physics, Imperial College London , London, SW7 2AZ, United Kingdom }}
\newcommand{\AFFkek}{\affiliation{High Energy Accelerator Research Organization (KEK), Tsukuba, Ibaraki 305-0801, Japan }}
\newcommand{\AFFkobe}{\affiliation{Department of Physics, Kobe University, Kobe, Hyogo 657-8501, Japan}}
\newcommand{\AFFkyoto}{\affiliation{Department of Physics, Kyoto University, Kyoto, Kyoto 606-8502, Japan}}
\newcommand{\AFFliv}{\affiliation{Department of Physics, University of Liverpool, Liverpool, L69 7ZE, United Kingdom}}
\newcommand{\AFFmiyagi}{\affiliation{Department of Physics, Miyagi University of Education, Sendai, Miyagi 980-0845, Japan}}
\newcommand{\AFFnagoya}{\affiliation{Institute for Space-Earth Environmental Research, Nagoya University, Nagoya, Aichi 464-8602, Japan}}
\newcommand{\AFFkmi}{\affiliation{Kobayashi-Maskawa Institute for the Origin of Particles and the Universe, Nagoya University, Nagoya, Aichi 464-8602, Japan}}
\newcommand{\AFFpol}{\affiliation{National Centre For Nuclear Research, 02-093 Warsaw, Poland}}
\newcommand{\AFFsuny}{\affiliation{Department of Physics and Astronomy, State University of New York at Stony Brook, NY 11794-3800, USA}}
\newcommand{\AFFokayama}{\affiliation{Department of Physics, Okayama University, Okayama, Okayama 700-8530, Japan }}
\newcommand{\AFFosaka}{\affiliation{Department of Physics, Osaka University, Toyonaka, Osaka 560-0043, Japan}}
\newcommand{\AFFox}{\affiliation{Department of Physics, Oxford University, Oxford, OX1 3PU, United Kingdom}}
\newcommand{\AFFqmul}{\affiliation{School of Physics and Astronomy, Queen Mary University of London, London, E1 4NS, United Kingdom}}
\newcommand{\AFFregina}{\affiliation{Department of Physics, University of Regina, 3737 Wascana Parkway, Regina, SK, S4SOA2, Canada}}
\newcommand{\AFFseoul}{\affiliation{Department of Physics and Astronomy, Seoul National University, Seoul 151-742, Korea}}
\newcommand{\AFFsheff}{\affiliation{School of Mathematical and Physical Sciences, University of Sheffield, S3 7RH, Sheffield, United Kingdom}}
\newcommand{\AFFshizuokasc}{\affiliation{Department of Informatics in
Social Welfare, Shizuoka University of Welfare, Yaizu, Shizuoka, 425-8611, Japan}}
\newcommand{\AFFstfc}{\affiliation{STFC, Rutherford Appleton Laboratory, Harwell Oxford, and Daresbury Laboratory, Warrington, OX11 0QX, United Kingdom}}
\newcommand{\AFFskk}{\affiliation{Department of Physics, Sungkyunkwan University, Suwon 440-746, Korea}}
\newcommand{\AFFtodai}{\affiliation{Department of Physics, University of Tokyo, Bunkyo, Tokyo 113-0033, Japan }}
\newcommand{\AFFtit}{\affiliation{Department of Physics, Institute of Science Tokyo, Meguro, Tokyo 152-8551, Japan }}
\newcommand{\AFFtus}{\affiliation{Department of Physics and Astronomy, Faculty of Science and Technology, Tokyo University of Science, Noda, Chiba 278-8510, Japan }}
\newcommand{\AFFtriumf}{\affiliation{TRIUMF, 4004 Wesbrook Mall, Vancouver, BC, V6T2A3, Canada }}
\newcommand{\AFFtokai}{\affiliation{Department of Physics, Tokai University, Hiratsuka, Kanagawa 259-1292, Japan}}
\newcommand{\AFFtsinghua}{\affiliation{Department of Engineering Physics, Tsinghua University, Beijing, 100084, China}}
\newcommand{\AFFynu}{\affiliation{Department of Physics, Yokohama National University, Yokohama, Kanagawa, 240-8501, Japan}}
\newcommand{\AFFllr}{\affiliation{Ecole Polytechnique, IN2P3-CNRS, Laboratoire Leprince-Ringuet, F-91120 Palaiseau, France }}
\newcommand{\AFFbari}{\affiliation{ Dipartimento Interuniversitario di Fisica, INFN Sezione di Bari and Universit\`a e Politecnico di Bari, I-70125, Bari, Italy}}
\newcommand{\AFFnapoli}{\affiliation{Dipartimento di Fisica, INFN Sezione di Napoli and Universit\`a di Napoli, I-80126, Napoli, Italy}}
\newcommand{\AFFroma}{\affiliation{INFN Sezione di Roma and Universit\`a di Roma ``La Sapienza'', I-00185, Roma, Italy}}
\newcommand{\AFFpadova}{\affiliation{Dipartimento di Fisica, INFN Sezione di Padova and Universit\`a di Padova, I-35131, Padova, Italy}}
\newcommand{\AFFkeio}{\affiliation{Department of Physics, Keio University, Yokohama, Kanagawa, 223-8522, Japan}}
\newcommand{\AFFwinnipeg}{\affiliation{Department of Physics, University of Winnipeg, MB R3J 3L8, Canada }}
\newcommand{\AFFkcl}{\affiliation{Department of Physics, King's College London, London, WC2R 2LS, UK }}
\newcommand{\AFFwarwick}{\affiliation{Department of Physics, University of Warwick, Coventry, CV4 7AL, UK }}
\newcommand{\AFFral}{\affiliation{Rutherford Appleton Laboratory, Harwell, Oxford, OX11 0QX, UK }}
\newcommand{\AFFwu}{\affiliation{Faculty of Physics, University of Warsaw, Warsaw, 02-093, Poland }}
\newcommand{\AFFbcit}{\affiliation{Department of Physics, British Columbia Institute of Technology, Burnaby, BC, V5G 3H2, Canada }}
\newcommand{\AFFtohoku}{\affiliation{Department of Physics, Faculty of Science, Tohoku University, Sendai, Miyagi, 980-8578, Japan }}
\newcommand{\AFFicise}{\affiliation{Institute For Interdisciplinary Research in Science and Education, ICISE, Quy Nhon, 55121, Vietnam }}
\newcommand{\AFFilance}{\affiliation{ILANCE, CNRS - University of Tokyo International Research Laboratory, Kashiwa, Chiba 277-8582, Japan}}
\newcommand{\AFFibs}{\affiliation{Center for Underground Physics, Institute for Basic Science (IBS), Daejeon, 34126, Korea}}
\newcommand{\AFFglasgow}{\affiliation{School of Physics and Astronomy, University of Glasgow, Glasgow, Scotland, G12 8QQ, United Kingdom}}
\newcommand{\AFFoecu}{\affiliation{Media Communication Center, Osaka Electro-Communication University, Neyagawa, Osaka, 572-8530, Japan}}
\newcommand{\AFFminn}{\affiliation{School of Physics and Astronomy, University of Minnesota, Minneapolis, MN  55455, USA}}
\newcommand{\AFFsilesia}{\affiliation{August Che\l{}kowski Institute of Physics, University of Silesia in Katowice, 75 Pu\l{}ku Piechoty 1, 41-500 Chorz\'{o}w, Poland}}
\newcommand{\AFFtoyama}{\affiliation{Faculty of Science, University of Toyama, Toyama City, Toyama 930-8555, Japan}}
\newcommand{\AFFbmcc}{\affiliation{Science Department, Borough of Manhattan Community College / City University of New York, New York, New York, 1007, USA.}}
\newcommand{\AFFnumazu}{\affiliation{National Institute of Technology, Numazu College, Numazu, Shizuoka  410-8501, Japan}}

\AFFicrr
\AFFkashiwa
\AFFmad
\AFFbmcc
\AFFbu
\AFFbcit
\AFFuci
\AFFcsu
\AFFcnm
\AFFduke
\AFFllr
\AFFgifu
\AFFgist
\AFFglasgow
\AFFuh
\AFFibs
\AFFicise
\AFFicl
\AFFbari
\AFFnapoli
\AFFpadova
\AFFroma
\AFFilance
\AFFkeio
\AFFkek
\AFFkcl
\AFFkobe
\AFFkyoto
\AFFliv
\AFFminn
\AFFmiyagi
\AFFnagoya
\AFFkmi
\AFFpol
\AFFnumazu
\AFFsuny
\AFFokayama
\AFFoecu
\AFFox
\AFFral
\AFFseoul
\AFFsheff
\AFFshizuokasc
\AFFsilesia
\AFFstfc
\AFFskk
\AFFtohoku
%\AFFtokai
%\AFFtokyo
\AFFtodai
\AFFipmu
\AFFtit
\AFFtus
\AFFtoyama
\AFFtriumf
\AFFtsinghua
\AFFwu
\AFFwarwick
\AFFwinnipeg
\AFFynu

%%%%%%%%%%%%%%%%%%%%%%%%%%%%%%%%%%%%%%%%%%%%%%%%%%%%%%%%%%%%%%%%%%%%
%ICRR
\author[0000-0002-2110-5130]{K.~Abe}
\AFFicrr
\AFFipmu
\author[0000-0001-6440-933X]{Y.~Asaoka}
\AFFicrr
\AFFipmu
\author[0000-0003-3273-946X]{M.~Harada}
\AFFicrr
\author[0000-0002-8683-5038]{Y.~Hayato}
\AFFicrr
\AFFipmu
\author[0000-0003-1229-9452]{K.~Hiraide}
\AFFicrr
\AFFipmu
\author[0000-0002-8766-3629]{K.~Hosokawa}
\AFFicrr
\AFFipmu
\author{T.~H.~Hung}
\AFFicrr
\author[0000-0002-7791-5044]{K.~Ieki}
\author[0000-0002-4177-5828]{M.~Ikeda}
\AFFicrr
\AFFipmu
\author{J.~Kameda}
\AFFicrr
\AFFipmu
% ...

%%%%%%%%%%%%%%%%%%%%%%%%%%%%%%%%%%%%%%%%%%%%%%%%%%%%%%%%%%%%%%%%%%%%%

% \collaboration{The Super-Kamiokande Collaboration}
\collaboration{2000}{The Super-Kamiokande Collaboration}
\noaffiliation

%% file: SK-paper-acknowledgements-20251121.tex
%Paper Acknowledgements(as of Nov. 22, 2024)

%Short list (in case number of lines are limited in short letter papers)

We gratefully acknowledge cooperation of the Kamioka Mining and Smelting Company.
The Super-Kamiokande experiment was built and has been operated with funding from the
Japanese Ministry of Education, Science, Sports and Culture, 
and the U.S. Department of Energy.

%Full list (for full papers and longer papers)

We gratefully acknowledge the cooperation of the Kamioka Mining and Smelting Company. The Super-Kamiokande experiment has been built and operated from funding by the Japanese Ministry of Education, Culture, Sports, Science and Technology; the U.S. Department of Energy; and the U.S. National Science Foundation. Some of us have been supported by funds from the National Research Foundation of Korea (NRF-2009-0083526, NRF-2022R1A5A1030700, NRF-2022R1A3B1078756, RS-2025-00514948) funded by the Ministry of Science, Information and Communication Technology (ICT); the Institute for Basic Science (IBS-R016-Y2); and the Ministry of Education (2018R1D1A1B07049158, 2021R1I1A1A01042256, RS-2024-00442775); the Japan Society for the Promotion of Science; the National Natural Science Foundation of China (Grants No. 12375100 and 12521007); the Spanish Ministry of Science, Universities and Innovation (grant PID2021-124050NB-C31); the Natural Sciences and Engineering Research Council (NSERC) of Canada; the Scinet and Digital Research of Alliance Canada; the National Science Centre (UMO-2018/30/E/ST2/00441 and UMO-2022/46/E/ST2/00336) and the Ministry of  Science and Higher Education (2023/WK/04), Poland; the Science and Technology Facilities Council (STFC) and Grid for Particle Physics (GridPP), UK; the European Union’s Horizon 2020 Research and Innovation Programme H2020-MSCA-RISE-2018 JENNIFER2 grant agreement no.822070, H2020-MSCA-RISE-2019 SK2HK grant agreement no. 872549; and European Union's Next Generation EU/PRTR  grant CA3/RSUE2021-00559; the National Institute for Nuclear Physics (INFN), Italy.